\UseRawInputEncoding
\documentclass[twocolumn]{aastex62}
\usepackage{amsmath,amstext}
\usepackage[T1]{fontenc}
\usepackage[figure,figure*]{hypcap}
\usepackage{graphicx}
\usepackage{epstopdf}
\usepackage{gensymb}
\usepackage{subfigure}

\graphicspath{{./}{figures/}}
\received{\today}
\revised{\today}
\accepted{xx xx, xxxx}

\newcommand{\msun}{$\mathrm{M_{\odot}}$}
\newcommand{\Ni}{$^{56}$Ni\ }

\shorttitle{\mbox{AT2018cow}}
\shortauthors{Xiang et al.}

\begin{document}

\title{The Peculiar Transient AT2018cow: A Possible Origin of A Type Ibn/IIn Supernova}

\author{Danfeng Xiang}
\affiliation{Physics Department and Tsinghua Center for Astrophysics (THCA), Tsinghua University, Beijing, 100084, China}

\author{Xiaofeng Wang}\thanks{E-mail: wang\_xf@mail.tsinghua.edu.cn}
\affiliation{Physics Department and Tsinghua Center for Astrophysics (THCA), Tsinghua University, Beijing, 100084, China}
\affiliation{Beijing Planetarium, Beijing Academy of Sciences and Technology, Beijing, 100044, China}
\affiliation{Purple Mountain Observatory, Chinese Academy of Sciences, Nanjing, 210023, China}

\author{Weili Lin}
\author{Jun Mo}
\author{Han Lin}
\affiliation{Physics Department and Tsinghua Center for Astrophysics (THCA), Tsinghua University, Beijing, 100084, China}

\author{Jamison Burke}
\affiliation{Las Cumbres Observatory, 6740 Cortona Drive, Suite 102, Goleta, CA 93117-5575, USA}
\affiliation{Department of Physics, University of California, Santa Barbara, CA 93106-9530, USA}
\author{Daichi Hiramatsu}
\affiliation{Las Cumbres Observatory, 6740 Cortona Drive, Suite 102, Goleta, CA 93117-5575, USA}
\affiliation{Department of Physics, University of California, Santa Barbara, CA 93106-9530, USA}
\author{Griffin Hosseinzadeh}
\affiliation{Center for Astrophysics \textbar{} Harvard \& Smithsonian, 60 Garden Street, Cambridge, MA 02138-1516, USA}
\author{D. Andrew Howell}
\affiliation{Las Cumbres Observatory, 6740 Cortona Drive, Suite 102, Goleta, CA 93117-5575, USA}
\affiliation{Department of Physics, University of California, Santa Barbara, CA 93106-9530, USA}
\author{Curtis McCully}
\affiliation{Las Cumbres Observatory, 6740 Cortona Drive, Suite 102, Goleta, CA 93117-5575, USA}
\affiliation{Department of Physics, University of California, Santa Barbara, CA 93106-9530, USA}
\author{Stefan Valenti}
\affiliation{Department of Physics and Astronomy, University of California, 1 Shields Avenue, Davis, CA 95616-5270, USA}

\author{J\'{o}zsef Vink\'{o}}
\affiliation{Konkoly Observatory, Research Centre for Astronomy and Earth Sciences,
Konkoly-Thege M. \'{u}t 15-17, Budapest 1121, Hungary}
\affiliation{ELTE E\"otv\"os Lor\'and University, Institute of Physics,
P\'azmany P\'eter s\'et\'any 1/A, Budapest, 1117, Hungary}
\affiliation{Department of Optics \& Quantum Electronics, University of Szeged,
D\'om t\'er 9, Szeged, 6720 Hungary}
\affiliation{Department of Astronomy, University of Texas at Austin, Austin, TX, 78712, USA}
\author{J. Craig Wheeler}
\affiliation{Department of Astronomy, University of Texas at Austin, Austin, TX, 78712, USA}

\author{Shuhrat A. Ehgamberdiev}
\affiliation{Ulugh Beg Astronomical Institute, Uzbekistan Academy of Sciences, Uzbekistan, Tashkent, 100052, Uzbekistan}
\author{Davron Mirzaqulov}
\affiliation{Ulugh Beg Astronomical Institute, Uzbekistan Academy of Sciences, Uzbekistan, Tashkent, 100052, Uzbekistan}

\author{Attila B\'odi}
\affiliation{Konkoly Observatory, Research Centre for Astronomy and Earth Sciences,
Konkoly-Thege M. \'{u}t 15-17, Budapest 1121, Hungary}
\affiliation{CSFK Lend\"ulet Near-Field Cosmology Research Group}
\affiliation{ELTE E\"otv\"os Lor\'and University, Institute of Physics,
P\'azmany P\'eter s\'et\'any 1/A, Budapest, 1117, Hungary}
\author{Zs\'ofia Bogn\'ar}
\affiliation{Konkoly Observatory, Research Centre for Astronomy and Earth Sciences,
Konkoly-Thege M. \'{u}t 15-17, Budapest 1121, Hungary}
\affiliation{CSFK Lend\"ulet Near-Field Cosmology Research Group}
\affiliation{ELTE E\"otv\"os Lor\'and University, Institute of Physics,
P\'azmany P\'eter s\'et\'any 1/A, Budapest, 1117, Hungary}
\author{Borb\'ala Cseh}
\affiliation{Konkoly Observatory, Research Centre for Astronomy and Earth Sciences,
Konkoly-Thege M. \'{u}t 15-17, Budapest 1121, Hungary}
\author{Ott\'o Hanyecz}
\affiliation{Konkoly Observatory, Research Centre for Astronomy and Earth Sciences,
Konkoly-Thege M. \'{u}t 15-17, Budapest 1121, Hungary}
\author{Bernadett Ign\'acz}
\affiliation{Konkoly Observatory, Research Centre for Astronomy and Earth Sciences,
Konkoly-Thege M. \'{u}t 15-17, Budapest 1121, Hungary}
\author{Csilla Kalup}
\affiliation{Konkoly Observatory, Research Centre for Astronomy and Earth Sciences,
Konkoly-Thege M. \'{u}t 15-17, Budapest 1121, Hungary}
\author{R\'eka K\"onyves-T\'oth}
\affiliation{Konkoly Observatory, Research Centre for Astronomy and Earth Sciences,
Konkoly-Thege M. \'{u}t 15-17, Budapest 1121, Hungary}
\author{Levente Kriskovics}
\affiliation{Konkoly Observatory, Research Centre for Astronomy and Earth Sciences,
Konkoly-Thege M. \'{u}t 15-17, Budapest 1121, Hungary}
\affiliation{E\"otv\"os Lor\'and University, Institute of Physics,
P\'azmany P\'eter s\'et\'any 1/A, Budapest, 1117, Hungary}
\author{Andr\'{a}s Ordasi}
\affiliation{Konkoly Observatory, Research Centre for Astronomy and Earth Sciences,
Konkoly-Thege M. \'{u}t 15-17, Budapest 1121, Hungary}
\author{Andr\'as P\'al}
\affiliation{Konkoly Observatory, Research Centre for Astronomy and Earth Sciences,
Konkoly-Thege M. \'{u}t 15-17, Budapest 1121, Hungary}
\affiliation{E\"otv\"os Lor\'and University, Institute of Physics,
P\'azmany P\'eter s\'et\'any 1/A, Budapest, 1117, Hungary}
\affiliation{E\"otv\"os Lor\'and University, Department of Astronomy,
P\'azmany P\'eter s\'et\'any 1/A, Budapest, 1117, Hungary}
\author{Kriszti\'an S\'arneczky}
\affiliation{Konkoly Observatory, Research Centre for Astronomy and Earth Sciences,
Konkoly-Thege M. \'{u}t 15-17, Budapest 1121, Hungary}
\author{B\'alint Seli}
\affiliation{Konkoly Observatory, Research Centre for Astronomy and Earth Sciences,
Konkoly-Thege M. \'{u}t 15-17, Budapest 1121, Hungary}
\author{R\'obert Szak\'ats}
\affiliation{Konkoly Observatory, Research Centre for Astronomy and Earth Sciences,
Konkoly-Thege M. \'{u}t 15-17, Budapest 1121, Hungary}

\author{T. Arranz-Heras}
\affiliation{Observadores de Supernovas Group (ObSN), Spain}
\affiliation{Obs. Las Pequeras, 40470 Navas de Oro, Segovia, Spain}
\author{R. Benavides-Palencia}
\affiliation{Observadores de Supernovas Group (ObSN), Spain}
\affiliation{Obs. Posadas MPC J53, 14730 Posadas, C\'{o}rdoba, Spain}
\author{D. Cejudo-Mart\'{i}nez}
\affiliation{Observadores de Supernovas Group (ObSN), Spain}
\affiliation{Obs. El Gallinero, 28192 El Berrueco, Madrid, Spain}
\author{P. De la Fuente-Fern\'{a}ndez}
\affiliation{Observadores de Supernovas Group (ObSN), Spain}
\affiliation{Obs. Llanes, 33784 Llanes, Asturias, Spain}
\author{A. Escart\'{i}n-P\'{e}rez}
\affiliation{Observadores de Supernovas Group (ObSN), Spain}
\affiliation{Obs. Belako, 48100 Mungia, Vizcaya, Spain}
\author{F. Garc\'{i}a-De la Cuesta}
\affiliation{Observadores de Supernovas Group (ObSN), Spain}
\affiliation{Obs. La Vara MPC J38, 33784 Mu\~{n}as de Arriba, Asturias, Spain}
\author{J.L. Gonz\'{a}lez-Carballo}
\affiliation{Observadores de Supernovas Group (ObSN), Spain}
\affiliation{Obs. Cerro del Viento MPC I84, 06010 Badajoz, Spain}
\author{R. Gonz\'{a}lez-Farf\'{a}n}
\affiliation{Observadores de Supernovas Group (ObSN), Spain}
\affiliation{Obs. Uraniborg MPC Z55, 41400 \'{E}cija, Sevilla, Spain}
\author{F. Lim\'{o}n-Mart\'{i}nez}
\affiliation{Observadores de Supernovas Group (ObSN), Spain}
\affiliation{Obs. Mazariegos MPC I99, 34170 Mazariegos, Palencia, Spain}
\author{A. Mantero}
\affiliation{Observadores de Supernovas Group (ObSN), Spain}
\affiliation{Obs. Bernezzo MPC C77, 12010 Bernezzo, Cuneo, Italy}
\author{R. Naves-Nogu\'{e}s}
\affiliation{Observadores de Supernovas Group (ObSN), Spain}
\affiliation{Obs. Montcabrer MPC 213, 08348 Cabrils, Barcelona, Spain}
\author{M. Morales-Aimar}
\affiliation{Observadores de Supernovas Group (ObSN), Spain}
\affiliation{Obs. Sencelles MPC K14, 07140 Sencelles, Islas Baleares, Spain}
\author{V. R. Ru\'{i}z-Ru\'{i}z}
\affiliation{Observadores de Supernovas Group (ObSN), Spain}
\affiliation{New Mexico Skies (USA), iTelescope Siding Spring (Asutralia), AstroCamp (Nerpio, Spain)}
\author{F.C. Sold\'{a}n-Alfaro}
\affiliation{Observadores de Supernovas Group (ObSN), Spain}
\affiliation{Obs. Amanecer de Arrakis MPC Z74, 41500 Alcal\'{a} de Guadaira, Sevilla, Spain}
\author{J. Valero-P\'{e}rez}
\affiliation{Observadores de Supernovas Group (ObSN), Spain}
\affiliation{Obs. Ponferrada MPC Z70, 24411 Ponferrada, Le\'{o}n, Spain}
\author{F. Violat-Bordonau}
\affiliation{Observadores de Supernovas Group (ObSN), Spain}
\affiliation{Obs. Norba Caesarina MPC Z71, 10195 C\'{a}ceres, Spain}
\author{Tianmeng Zhang}
\affiliation{Key Laboratory of Optical Astronomy, National Astronomical Observatories, Chinese Academy of Sciences, 10101, Beijing}
\affiliation{School of Astronomy and Space Science, University of Chinese Academy of Sciences, 101408, Beijing}
\author{Jujia Zhang}
\affiliation{Yunnan Observatories, Chinese Academy of Sciences, Kunming 650216, China}
\affiliation{Key Laboratory for the Structure and Evolution of Celestial Objects, Chinese Academy of Sciences, Kunming 650216, China}
\affiliation{Center for Astronomical Mega-Science, Chinese Academy of Sciences, 20A Datun Road, Chaoyang District, Beijing, 100012, China}
\author{Xue Li}
\author{Zhihao Chen}
\author{Hanna Sai}
\author{Wenxiong Li}
\affiliation{Physics Department and Tsinghua Center for Astrophysics (THCA), Tsinghua University, Beijing, 100084, China}

\begin{abstract}
We present our photometric and spectroscopic observations on the peculiar transient AT2018cow. The multi-band photometry covers from peak to $\sim$70 days and the spectroscopy ranges from 5 to $\sim$50 days. The rapid rise ($t_{\mathrm{r}}$$\lesssim$2.9 days), high luminosity ($M_{V,\mathrm{peak}}\sim-$20.8 mag) and fast decline after peak make AT2018cow stand out of any other optical transients. While we find that its light curves show high resemblance to those of type Ibn supernovae. Moreover, the spectral energy distribution remains high temperature of $\sim$14,000 K after $\sim$15 days since discovery. The spectra are featureless in the first 10 days, while some broad emission lines due to H, He, C and O emerge later, with velocity declining from $\sim14,000$ km s$^{-1}$ to $\sim$3000 km s$^{-1}$ at the end of our observations. Narrow and weak He I emission lines emerge in the spectra at $t>$20 days since discovery. These emission lines are reminiscent of the features seen in interacting supernovae like type Ibn and IIn subclasses. We fit the bolometric light curves with a model of circumstellar interaction (CSI) and radioactive decay (RD) of \Ni and find a good fit with ejecta mass $M_{\mathrm{ej}}\sim3.16$ \msun, circumstellar material mass $M_{\mathrm{CSM}}\sim0.04$ \msun, and ejected \Ni mass $M_{^{56}\mathrm{Ni}}\sim0.23$ \msun. The CSM shell might be formed in an eruptive mass ejection of the progenitor star. Furthermore, host environment of AT2018cow implies connection of AT2018cow with massive stars. Combining observational properties and the light curve fitting results, we conclude that AT2018cow might be a peculiar interacting supernova originated from a massive star.

\end{abstract}


\section{Introduction}\label{sec:intro}
The studies of time domain astronomy cover a variety of optical transients, including novae, supernovae (SNe), tidal disruption events (TDE), and kilonovae, etc.
With different physical origins, these transients exhibit a huge diversity in evolutionary properties, especially optical light curves.
The evolutionary time scales and luminosity of different transients are directly related to their physical origins.
There are a group of transients with very high luminosity and short timescale of evolution, such as the so-called fast evolving luminous transients (FELTs) \citep[e.g.][]{2018NatAs...2..307R}.
They have much faster rise and decline in light curves than regular SNe. And many of them have peak luminosity much higher than normal SNe, close to the superluminous supernovae (SLSNe, \citealp{2011Natur.474..487Q, 2017hsn..book..431H}).
The physical origins of these FELTs are still unclear.
Among them, some are characterized by very blue color, indicating high temperature, which are also called fast-rising blue optical transients (FBOTs, e.g. \citealp{2014ApJ...794...23D, 2016ApJ...819...35A}).

A recently discovered extragalactic transient, AT2018cow (ATLAS18qqn), has caught much attention due to its peculiar behaviour in its light curves and spectral evolution.
AT2018cow was discovered by ATLAS on MJD~58285.44 (UT Jun. 16.44, 2018, UT dates are used throughout this paper), with a magnitude of 14.76$\pm$0.10 mag in ATLAS {\it orange}-band \citep{2018ATel11727....1S}.
It is located far from the center of the host galaxy CGCG~137-068 ($z=0.0141$, $D_L=63$~Mpc\footnote{We assume a flat universe with $H_0=67.7~\mathrm{km~s^{-1}~{Mpc^{-1}}}$, $\Omega_{\mathrm{M}}=0.307$ \citep{2016A&A...594A..13P}.}).
This distance means that AT2018cow is as luminous as the peak of SNe Ia at discovery.
As soon as this transient source was reported, astronomers from all over the world were actively conducting its follow-up observations in all bands, including ultra-violet (UV), optical, X-ray, radio and $\gamma$-ray.
AT2018cow is found to evolve rapidly with a rise time less than 3 days and peak magnitude <$-$20 mag.
The photospheric temperature is measured to be $\sim$30,000 K near the peak and it still maintains high temperature of $\sim$15,000 K after $\sim$20 days after discovery \citep{2018ApJ...865L...3P, 2019MNRAS.484.1031P}.
All of these features suggest that AT2018cow can be put into the FBOTs.

The close distance makes AT2018cow the first FELT/FBOT which has well-sequenced photometric and spectroscopic observations in wavebands ranging from X-ray to radio (e.g., \citealp{2018ApJ...865L...3P, 2019MNRAS.484.1031P, 2019MNRAS.487.2505K, 2019ApJ...872...18M, 2019ApJ...871...73H}), making it a rare sample for the study of FBOT-like objects.
In previous studies, several possible physical mechanisms have been proposed for AT2018cow, e.g., tidal disruption of a star into an intermediate mass black hole (\citealp{2019MNRAS.484.1031P, 2019MNRAS.487.2505K}, Li et al. in prep.), central-engine powered supernova \citep{2018ApJ...865L...3P, 2019ApJ...872...18M}, interaction of a condensed CSM and the supernova shock \citep{2019ApJ...872...18M, 2020ApJ...903...66L}, electron-capture collapse of a white dwarf \citep{2019MNRAS.487.5618L}. And \cite{2019ApJ...872...18M} suggests that there should be a deeply embedded X-ray source in an asymmetrical ejecta.

In this paper, we present our optical photometric and spectroscopic observations of AT2018cow. Spectroscopic observations spanned from Jun. 21, 2018 to Aug. 14, 2018. and photometric observations lasted until September 21, 2018. In Sec. \ref{sec:observation} we describe our spectroscopic and photometric observations as well as data processing. In Sec. \ref{sec:obs-prop} we analyse the observational properties of AT2018cow, including light-curve and spectral evolution. The analysis of the host galaxy is presented in Sec. \ref{sec:host_env}. In Sec. \ref{sec:lc_model} we explore the possible physical origins of AT2018cow. Further discussion and final summary are given in Sec. \ref{sec:discussion} and \ref{sec:summary}, respectively.

\section{Observations and Data Reduction}\label{sec:observation}
\subsection{Photometric Observations}\label{subsec:phot}
The optical photometric observations of AT2018cow were monitored by several observatories, including the 0.8-m \mbox{Tsinghua} University-NAOC telescope \citep[TNT,][]{2012RAA....12.1585H} at Xinglong Observatory of NAOC, the AZT-22 \mbox{1.5-m} telescope (hereafter AZT) at Maidanak Astronomical \mbox{Observatory} \citep{2018NatAs...2..349E}, telescopes of the Las Cumbres Observatory network (LCO), and telescope of Konkoly Observatory in Hungary (hereafter KT). Photometric and spectroscopic data from LCO were obtained via the Global Supernova Project (GSP). We also collected early time photometric data from Observadores de Supernovas Group (ObSN) in Spain.
The TNT and LCO observations were obtained in standard Johnson-Cousin $UBV$ bands and SDSS $gri$ bands.
Long time and short-cadenced observations in $UBVRI$ bands were obtained by AZT. 
The Konkoly observations were obtained in $BVRI$ bands. Data from ObSN were obtained in $BVRI$ and $gr$ bands.
The entire dataset covers phases from MJD 58286.89 (Jun. 17.89, 2018) to MJD 58348.74 (Aug. 18.74, 2018).
The earliest photometric data point comes from ObSN in $V$-band on MJD 58286.89, which is $\sim$0.27 day earlier than that presented in \cite{2018ApJ...865L...3P}.
Besides the fast rise, the object faded very quickly. The late time photometry may be influenced by contamination from the galaxy. Thus for AZT, LCO and KT, we obtained reference images in each band in Mar. 2019, Oct. 2018, and Feb. 2019, respectively.
The reference images were obtained in all corresponding bands except for the $U$-band of AZT.
For TNT images, since the source is still bright during observations, the influence of the background is negligible.
Although the observations continued after Aug. 18, 2018, the object became too faint to be distinguished from the background.

\begin{figure}
\centering
\includegraphics[width=1.2\linewidth]{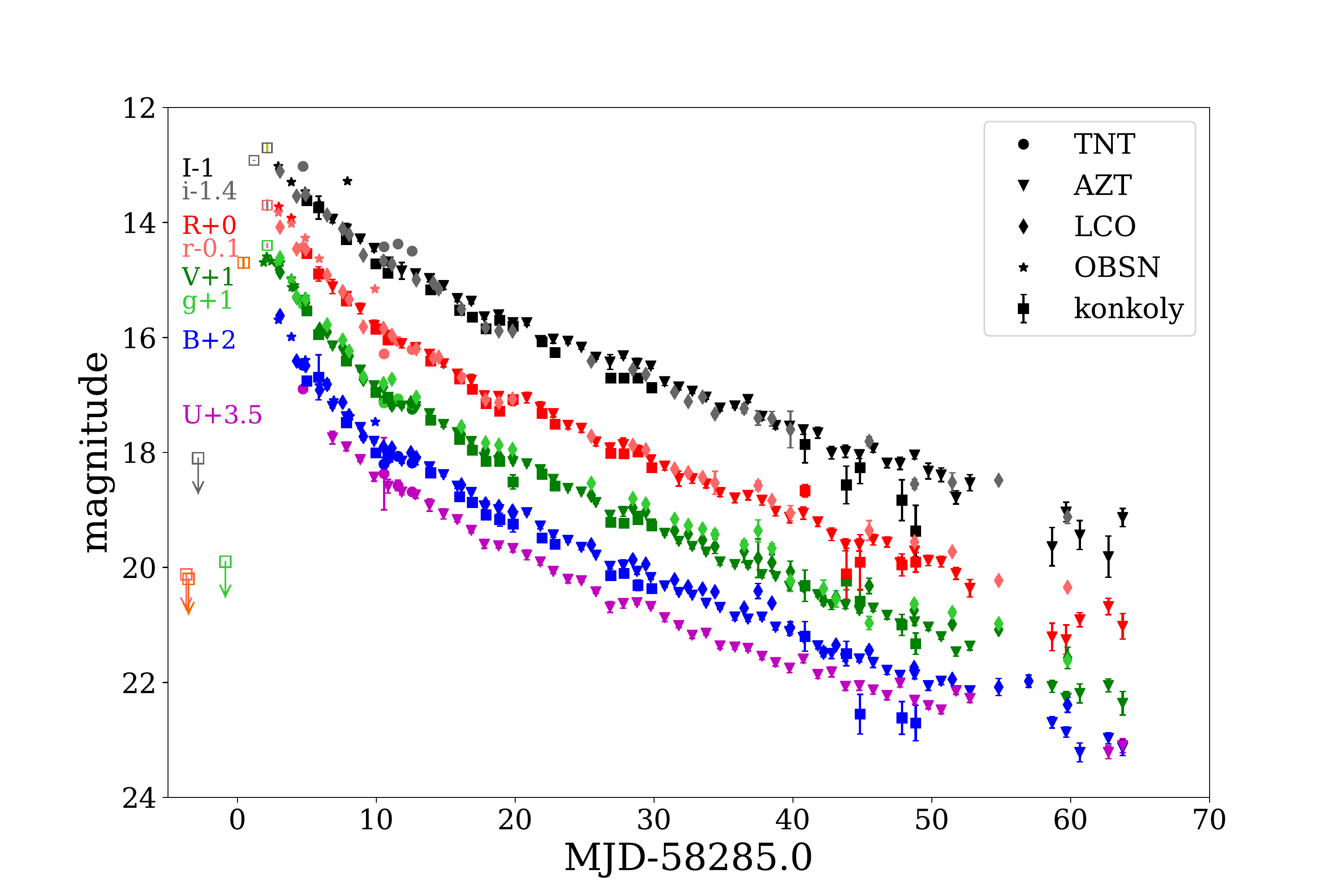}
\caption{Light curves obtained from various telescopes. Discovery magnitude in $orange$-band from \cite{2018ATel11727....1S} (orange) and early follow-up photometry from \cite{2018ATel11738....1F} and \cite{2018ApJ...865L...3P} are also plotted as empty squares. The pre-discovery detection limits are from \cite{2018ATel11738....1F} and \cite{2018ApJ...865L...3P}. The magnitudes in different bands are shifted for better display. \label{fig:lc}}
\end{figure}

All $UBVRI$ and $gri$ images are pre-processed using standard \textsc{IRAF}\footnote{IRAF is distributed by the National Optical Astronomy Observatories, which are operated by the Association of Universities for Research in Astronomy, Inc., under cooperative agreement with the National Science Foundation (NSF).} routines, which includes corrections for bias, flat field, and removal of cosmic rays.
To remove the contamination from the host galaxy, we applied template subtraction to the AZT, LCO and KT images. Note that the $U$-band images were not host subtracted.
The instrumental magnitudes of both AT2018cow and the reference stars were then measured using the standard point spread function (PSF).
And then the instrumental magnitudes were converted to standard Johnson and SDSS gri-band magnitudes using the zero points and color terms of each telescope.
The resultant magnitudes are listed in Tab.~\ref{tab:lc}.
We also include the early photometry from \cite{2018ApJ...865L...3P} for comparison.
The light curves are shown in Fig.~\ref{fig:lc}.
\startlongtable
\begin{deluxetable}{lccccl}
\centering
\tablecaption{Portion of optical photometric observations of AT2018cow.\label{tab:lc}}
\tablehead{\colhead{MJD} &\colhead{mag.} &\colhead{mag. error} &\colhead{band} &\colhead{Telescope/reference}}
\startdata
58285.4400 &  14.700 &   0.100 &   o &       \cite{2018ATel11727....1S}\\
58286.1950 &  14.320 &   0.010 &   i &       \cite{2018ATel11738....1F}\\
58286.8880 &  13.695 &     ... &   V &            ObSN\\
58287.1130 &  13.593 &     ... &   V &            ObSN\\
58287.1500 &  13.400 &   0.050 &   g &   \cite{2018ApJ...865L...3P}\\
58287.1500 &  13.800 &   0.100 &   r &   \cite{2018ApJ...865L...3P}\\
58287.1500 &  14.100 &   0.100 &   i &   \cite{2018ApJ...865L...3P}\\
58287.4440 &  13.674 &     ... &   V &            ObSN\\
58287.9270 &  13.771 &     ... &   V &            ObSN\\
58287.9400 &  14.021 &     ... &   I &            ObSN\\
58287.9460 &  13.926 &     ... &   r &            ObSN\\
58287.9520 &  13.742 &     ... &   V &            ObSN\\
58287.9540 &  13.692 &     ... &   B &            ObSN\\
58287.9540 &  13.692 &     ... &   g &            ObSN\\
58287.9750 &  13.725 &     ... &   R &            ObSN\\
58288.0677 &  13.809 &   0.021 &   B &             LCO\\
58288.0677 &  13.939 &   0.013 &   V &             LCO\\
58288.0677 &  13.787 &   0.011 &   g &             LCO\\
58288.0677 &  14.573 &   0.016 &   i &             LCO\\
58288.0677 &  14.295 &   0.017 &   r &             LCO\\
\enddata
\end{deluxetable}

It can be seen that AT2018cow rises to a peak at MJD$\sim$58287.0 in $V$, $R$ and $I$ bands,  where the light curves are better sampled around the peak.
The latest non-detection limit is on MJD 58284.13 in $g$-band \citep{2018ApJ...865L...3P}, so the rise time of AT2018cow is less than 2.9 days. If we take the median of the first detection (i.e. discovery by ATLAS) MJD 58285.44 and the latest non-detection (i.e. MJD 58284.13) as the first light time, then the rise time is $\sim$2.2 days.
We apply an explosion time on MJD=58284.79$\pm$0.66 throughout this paper.
This rise time is too short compared to supernovae, which usually have rise time of more than 10 days.
After the peak, the light curves decline as fast as 0.33~mag~d$^{-1}$, 0.27~mag~d$^{-1}$, 0.22~mag~d$^{-1}$, within the first 10 days in $V$, $R$ and $I$-bands, respectively.

\subsection{Optical Spectroscopic Observations}\label{subsec:spec}
Our first spectrum was taken at Jul. 21, 2018 by the 2.16-m telescope at Xinglong Observatory of NAOC (hereafter XLT).
A total of 31 spectra were collected with different telescopes, including the XLT, the 2-m Faulkes Telescope North (FTN) of the Las Cumbres Observatory network, and the 9.2-m Hobby-Eberly Telescope (HET).
The details of the spectroscopic observations are listed in Tab.~\ref{tab:spec}.

\begin{figure}
\centering
\includegraphics[width=1.1\linewidth]{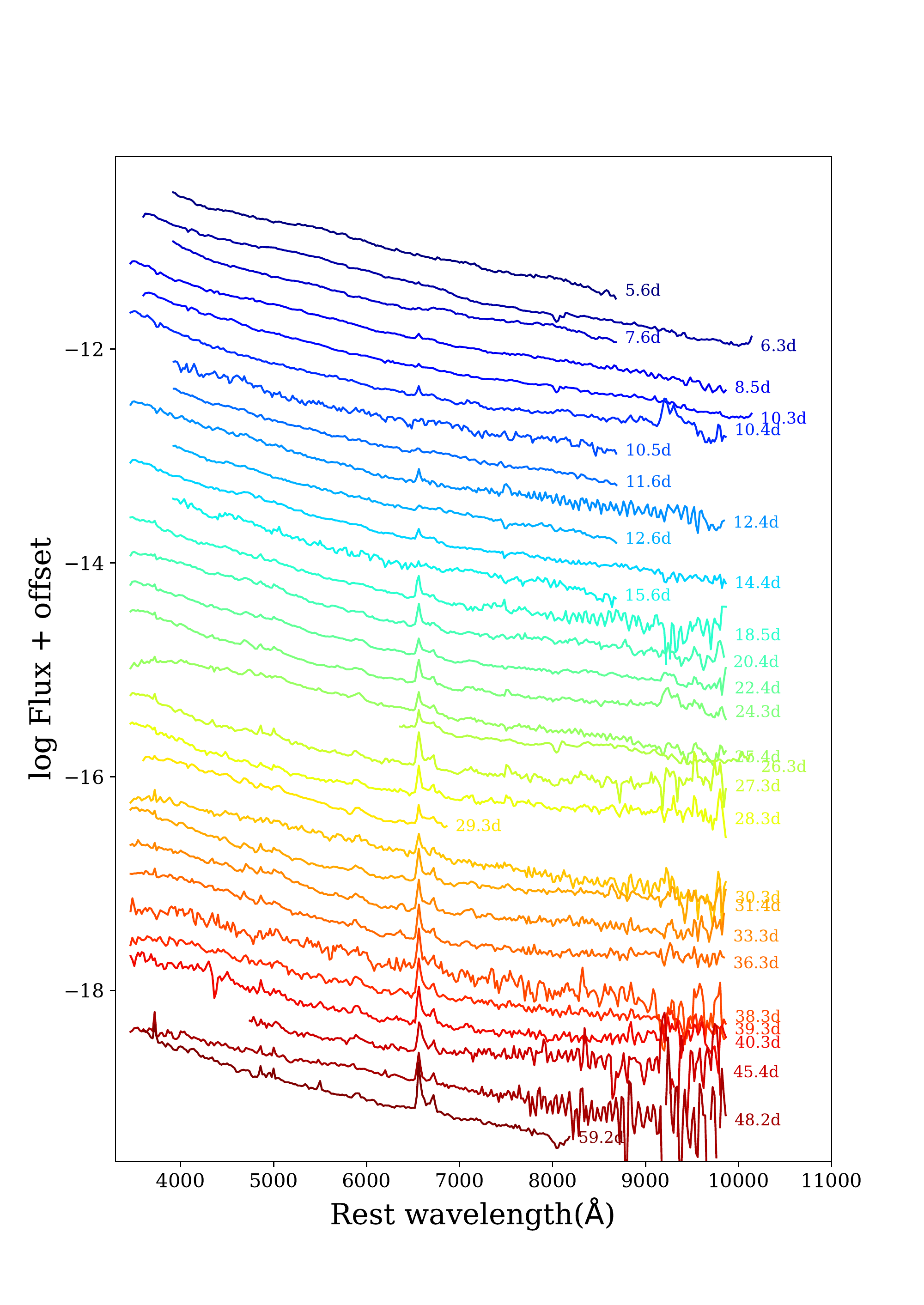}
\caption{Optical spectra of AT2018cow. The numbers indicate days since MJD 58285. Host galaxy emissions are not removed. The data are smoothed by a bin of 20~\AA\ for better display. \label{fig:spectra}}
\end{figure}

All spectra were reduced using the standard IRAF routines, which involves corrections for bias, flat field, and removal of cosmic rays.
The Fe/Ar and Fe/Ne arc lamp spectra obtained during the observation nights are used to calibrate the wavelength of the spectra, and standard stars observed on the same night at similar airmasses as the supernova were used to calibrate the flux of spectra.
The spectra were further corrected for continuum atmospheric extinction during flux calibration using mean extinction curves obtained at Xinglong Observatory and Haleakala Observatory in Hawaii, respectively.
Moreover, telluric lines were removed from the spectra of XLT and FTN.
We recalibrated the fluxes of the spectra to the multi-band photometry data. The UV data from \cite{2019MNRAS.484.1031P} are included in the recalibration process.
The recalibrated spectra are shown in Fig.~\ref{fig:spectra}.

On Sep. 17, 2019, when AT2018cow already faded away in the host galaxy, a spectrum was obtained at the site of AT2018cow by HET.
There are some narrow absorption lines in the resultant spectrum, which are an artifact of data reduction.
HET LRS2 is an IFU spectrograph having 280 individual fibers packed close together in a rectangular pattern, with a field-of-view of 12"$\times$6", which is smaller than the size of the host galaxy of AT2018cow.
Since the data reduction pipeline determines the background by combining the fibers having the lowest flux level, the background will necessarily contain some of the galaxy features. Thus the spectra show some fake absorption lines resulting from subtraction of the emission lines from other faint part of the host galaxy. These fake lines are manually removed from the spectrum.
A detailed analysis on this spectrum is presented in Sec.~\ref{sec:host_env}.
\startlongtable
\begin{deluxetable*}{lccccc}
\centering
\tablecaption{Log of optical spectroscopy of AT2018cow.\label{tab:spec}}
\tablehead{\colhead{UT} &\colhead{MJD} &\colhead{Telescope} &\colhead{Wav. range (\AA)} &\colhead{Instrument} &\colhead{Exposure time (s)}}
\startdata
2018/06/21.58  &58290.58  &XLT  &3970-8820  &BFOSC  &  2400\\
2018/06/22.32  &58291.32  &HET  &3640-10298  &LRS2  &   300\\
2018/06/23.64  &58292.64  &XLT  &3970-8820  &BFOSC  &  2400\\
2018/06/24.50  &58293.50  &FTN  &3500-10000  &FLOYDS  &  1200\\
2018/06/26.30  &58295.30  &HET  &3640-10300  &LRS2  &   500\\
2018/06/26.39  &58295.39  &FTN  &3500-10000  &FLOYDS  &  1200\\
2018/06/26.54  &58295.54  &XLT  &3970-8820  &BFOSC  &  1200\\
2018/06/27.57  &58296.57  &XLT  &3970-8820  &BFOSC  &  2400\\
2018/06/28.35  &58297.35  &FTN  &3500-10000  &FLOYDS  &  1200\\
2018/06/28.55  &58297.55  &XLT  &3970-8820  &BFOSC  &  1200\\
2018/06/30.38  &58299.38  &FTN  &3500-10000  &FLOYDS  &  1200\\
2018/07/01.57  &58300.57  &XLT  &3970-8820  &BFOSC  &  1500\\
2018/07/04.48  &58303.48  &FTN  &3500-10000  &FLOYDS  &  1200\\
2018/07/06.43  &58305.43  &FTN  &3500-10000  &FLOYDS  &  2700\\
2018/07/08.37  &58307.37  &FTN  &3500-10000  &FLOYDS  &  2700\\
2018/07/10.33  &58309.33  &FTN  &3500-10000  &FLOYDS  &  2700\\
2018/07/11.41  &58310.41  &FTN  &3500-10000  &FLOYDS  &  2700\\
2018/07/12.25  &58311.25  &HET  &6440-10300  &LRS2  &  1000\\
2018/07/13.32  &58312.32  &FTN  &3500-10000  &FLOYDS  &  2700\\
2018/07/14.35  &58313.35  &FTN  &3500-10000  &FLOYDS  &  2700\\
2018/07/15.25  &58314.25  &HET  &3640-6970  &LRS2  &   800\\
2018/07/16.31  &58315.31  &FTN  &3500-10000  &FLOYDS  &  2700\\
2018/07/17.35  &58316.35  &FTN  &3500-10000  &FLOYDS  &  2700\\
2018/07/19.35  &58318.35  &FTN  &3500-10000  &FLOYDS  &  3600\\
2018/07/22.28  &58321.28  &FTN  &3500-10000  &FLOYDS  &  3600\\
2018/07/24.34  &58323.34  &FTN  &3500-10000  &FLOYDS  &  3600\\
2018/07/25.32  &58324.32  &FTN  &3500-10000  &FLOYDS  &  3600\\
2018/07/26.34  &58325.34  &FTN  &3500-10000  &FLOYDS  &  3600\\
2018/07/31.37  &58330.37  &FTN  &4800-10000  &FLOYDS  &  3600\\
2018/08/03.25  &58333.25  &FTN  &3500-10000  &FLOYDS  &  3600\\
2018/08/14.18  &58344.18  &HET  &3640-8300  &LRS2  &  1800\\
2019/09/17.09  &58743.09  &HET  &3640-10200  &LRS2  &  1800\\
\enddata
\end{deluxetable*}

\section{Observational Properties}\label{sec:obs-prop}
\subsection{Light Curves and Color Evolution}\label{subsec:lc}
\begin{figure*}
\centering
\includegraphics[width=0.8\linewidth]{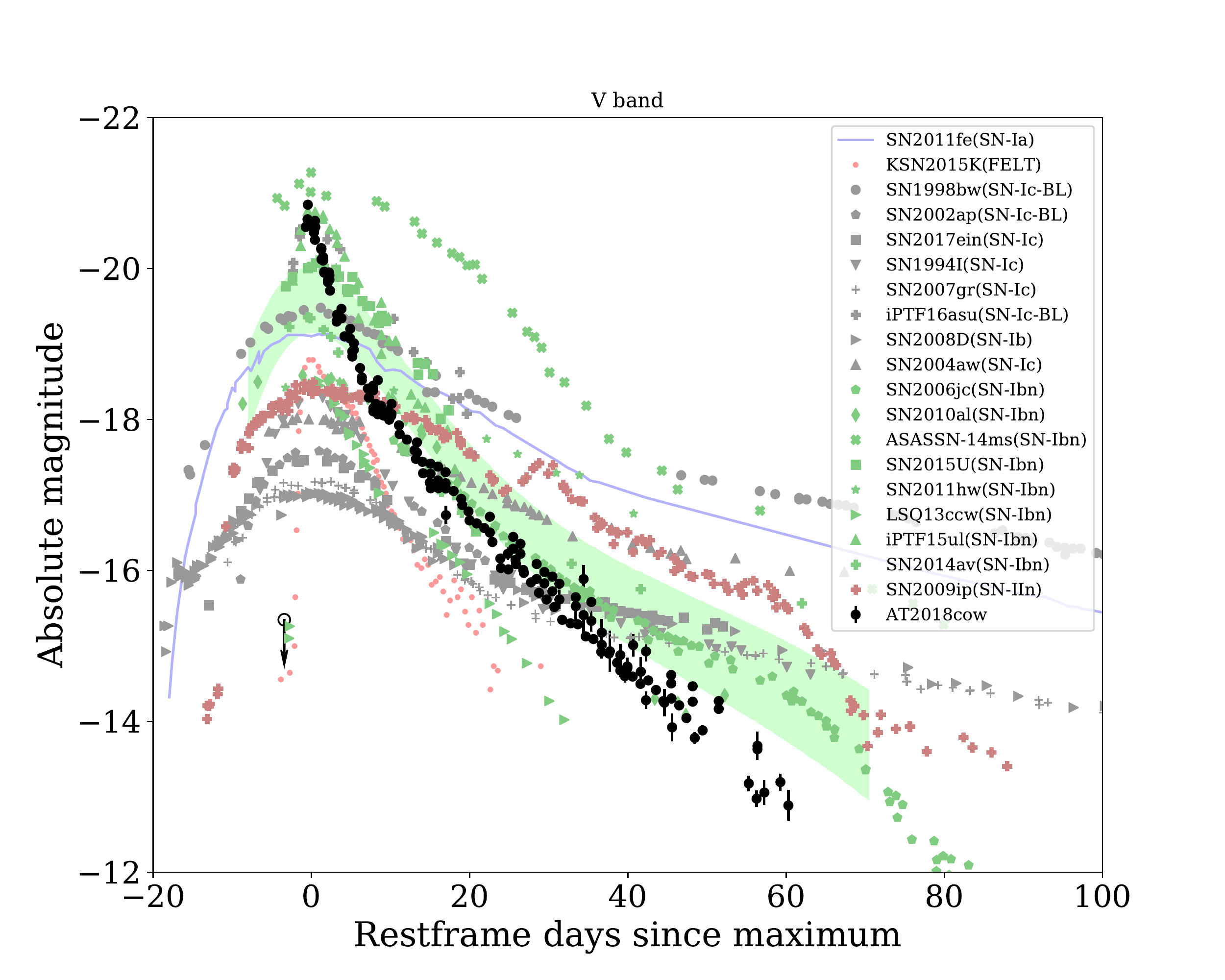}
\caption{The V-band light curves of AT2018cow compared with other optical transients. Different colors are used to distinguish object types. The green shaded area shows the template R-band light curves of SNe Ibn from \cite{2017ApJ...836..158H}. Data references: SN~2011fe \citep{2016ApJ...820...67Z}, KSN2015K \citep{2018NatAs...2..307R}, SN~1998bw \citep{1998Natur.395..670G, 2000ApJ...537L.127S, 1999PASP..111..964M}, SN~2002ap \citep{2003PASP..115.1220F}, SN~2017ein \citep{2019ApJ...871..176X, 2018ApJ...860...90V}, SN~1994I \citep{1996AJ....111..327R, 1994PASJ...46L.191Y}, SN~2007gr \citep{2014ApJ...790..120C}, iPTF16asu \citep{2017ApJ...851..107W},  SN~2008D \citep{2008Sci...321.1185M, 2009ApJ...702..226M, 2014ApJS..213...19B, 2014Ap&SS.354...89B}, SN~2004aw \citep{2006MNRAS.371.1459T}, SN~2006jc \citep{2014ApJS..213...19B, 2014Ap&SS.354...89B, 2011ApJ...741...97D},SN~2010al \citep{2017ApJS..233....6H, 2014Ap&SS.354...89B}, ASASSN-14ms (Wang, et al. 2020, in prep.), SN~2015U \citep{2016MNRAS.461.3057S, 2015IBVS.6140....1T}, SN~2011hw \citep{2014Ap&SS.354...89B, 2012MNRAS.426.1905S}, LSQ13ccw \citep{2015A&A...579A..40S}, iPTF15ul \citep{2017ApJ...836..158H}, SN~2014av \citep{2016MNRAS.456..853P}, SN~2009ip: \cite{2013MNRAS.430.1801M, 2015A&A...579A..40S}. Part of the reference data are obtained via the Open Supernova Catalog \citep{2017ApJ...835...64G}.\label{fig:lc_cmp_V}}
\end{figure*}

The light curves of AT2018cow show much faster evolution than other optical transients.
In Fig.~\ref{fig:lc_cmp_V} we compare the $V$-band light curves of AT2018cow with other SNe of different subtypes, including the peculiar fast-evolving transient KSN2015K \citep{2018NatAs...2..307R}.
One can see that both the rise and decline of AT2018cow are faster than any other known fast-evolving supernovae.
The rise time is very close to KSN2015K, while AT2018cow is about 2 mags brighter.
\replaced{For SLSNe, only the rapidly evolving object iPTF16asu \citep{2017ApJ...851..107W} is shown for comparison, since most SLSNe have much slower evolution.}{Most SLSNe have much slower evolution so we do not show them in the plot.}
AT2018cow is close to the type Ibn SN iPTF15ul \citep{2017ApJ...836..158H} in peak luminosity, while it is similar to the type Ibn SN~2006jc in terms of fast decline after the peak.
It should be noted that the high luminosity as well as the rapid evolution seen in AT 2018cow lie in the range of SNe~Ibn.

\begin{figure}[htb]
\centering
	\begin{minipage}{1.0\linewidth}
	\includegraphics[width=1.0\linewidth]{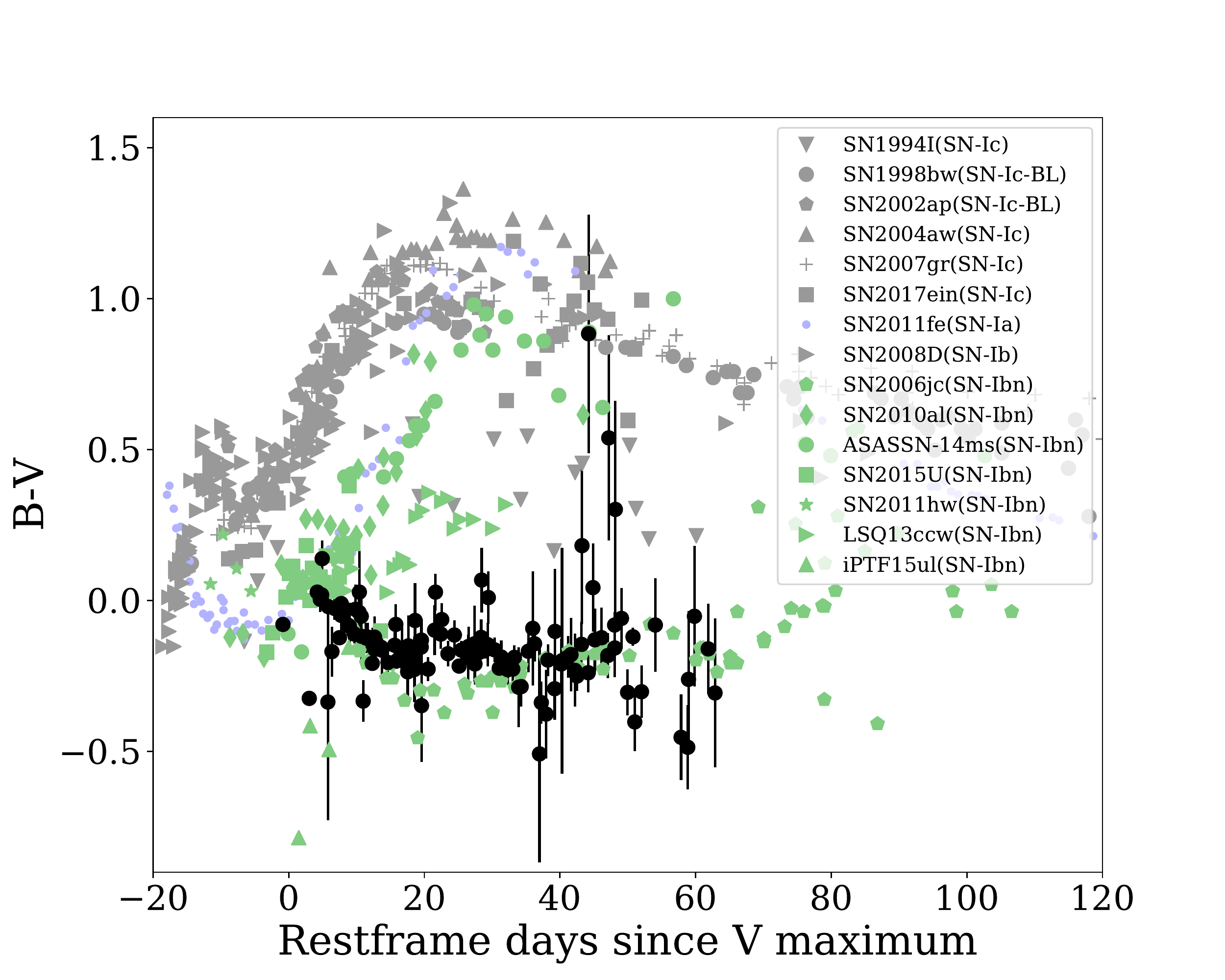}
	\end{minipage}
	\begin{minipage}{1.0\linewidth}
	\includegraphics[width=1.0\linewidth]{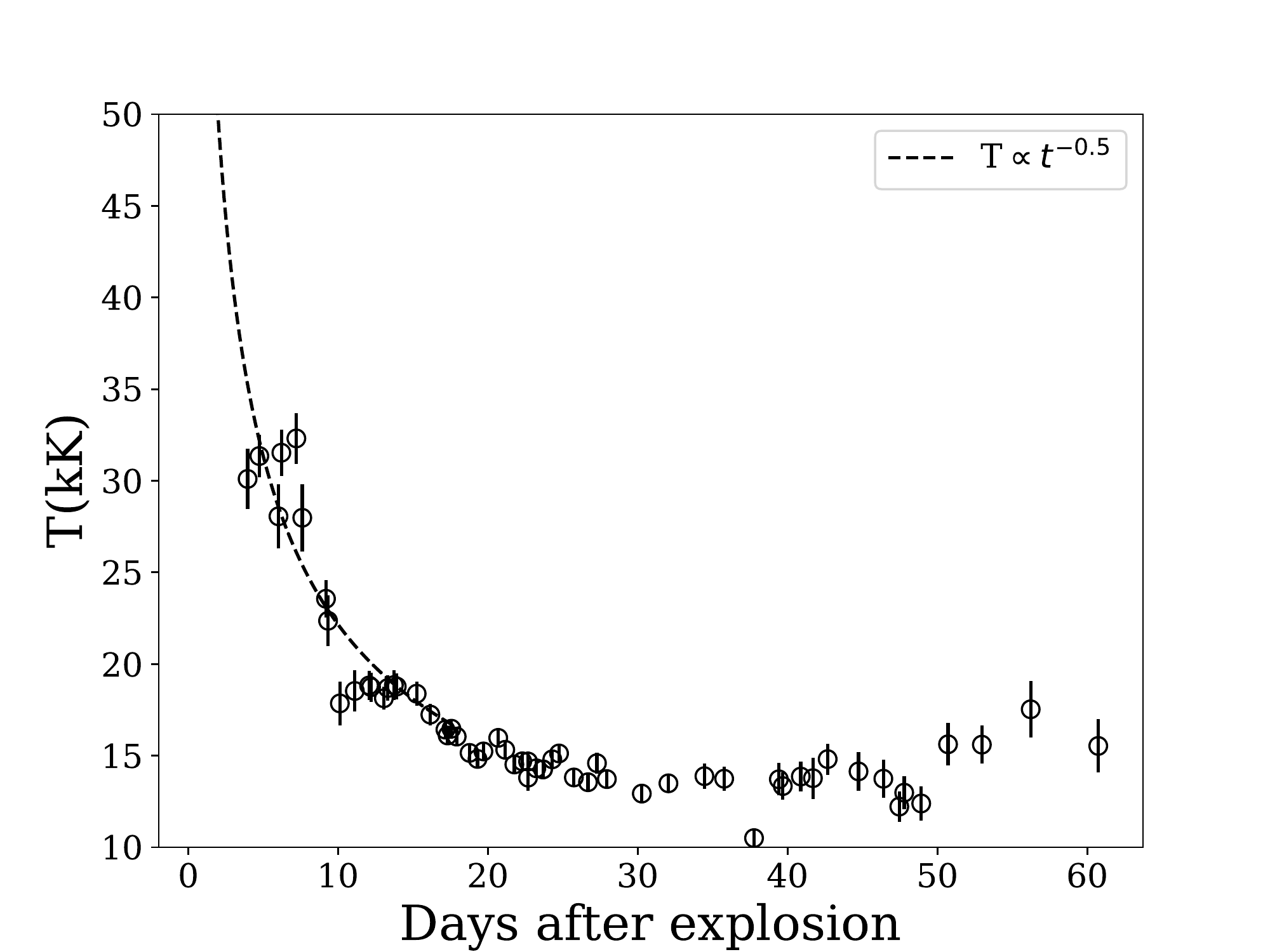}
	\end{minipage}
\caption{\textit{Upper: }$B-V$ evolution of AT2018cow, in comparison with other well observed SNe. Symbols and references are the same as in Fig.~\ref{fig:lc_cmp_V}. \textit{Lower}: temperature evolution of AT2018cow. The dased line shows the estimation of the early temperature as $T\propto t^{-0.5}$, which is used to estimate the early time bolometric luminosity (see Sec.~\ref{sec:lc_model}). \label{fig:B-V-color}}
\end{figure}

During our observations, AT2018cow maintains very blue color (i.~e. $B-V\sim-0.1$~mag, Fig.~\ref{fig:B-V-color}).
Thus, it should suffer little reddening from its host galaxy. This can also be verified by the absence of Na~I~D absorption line in the spectra.
We only consider the Galactic extinction of $E(B-V)$=0.08 \citep{2011ApJ...737..103S} for AT2018cow, and ignore the host extinction in this paper.
As also proposed by \cite{2019MNRAS.484.1031P}, the photospheric temperature of AT2018cow is as high as $\sim$30,000~K near the maximum light, and is still as high as $\sim$14,000~K at $\sim$50 days after discovery.
This is not seen in any other optical transients ever discovered.
For supernovae, the photospheric temperature can be high in early times but usually cools down to $\sim$5000~K in a few weeks after explosion, since the energy source is not strong enough to maintain a very high temperature.
So the color of normal SNe will become red in late phases.
In Fig.~\ref{fig:B-V-color} we show the $B-V$ color evolution of AT2018cow in comparison with other SNe.
The color evolution of AT2018cow resembles that of SN~2006jc.
Assuming a blackbody SED shape, the spectra of SN~2006jc also seem to present unusually high effective temperature, $\sim$15,000~K on day 8, then the temperature grows to 25,000~K on day 25 and drops to 15,000~K around day 60. The temperature decreases to $\sim$3,500 K and then keeps flat after day~80. Nevertheless, the interaction and blending of iron lines may indeed contribute to the high temperature.

Another interesting point is that the photospheric radius seems to be decreasing since the very beginning, unlike that of normal SNe, which will increase before peak and then decrease as a result of the expansion and dilution of the ejecta.
The absence of an expansion phase is the main problem of the supernova origin for AT2018cow.

\subsection{Spectral Evolution: Signatures of Interaction}\label{subsec:spectral_evolution}
\begin{figure}
\centering
\includegraphics[width=1.0\linewidth]{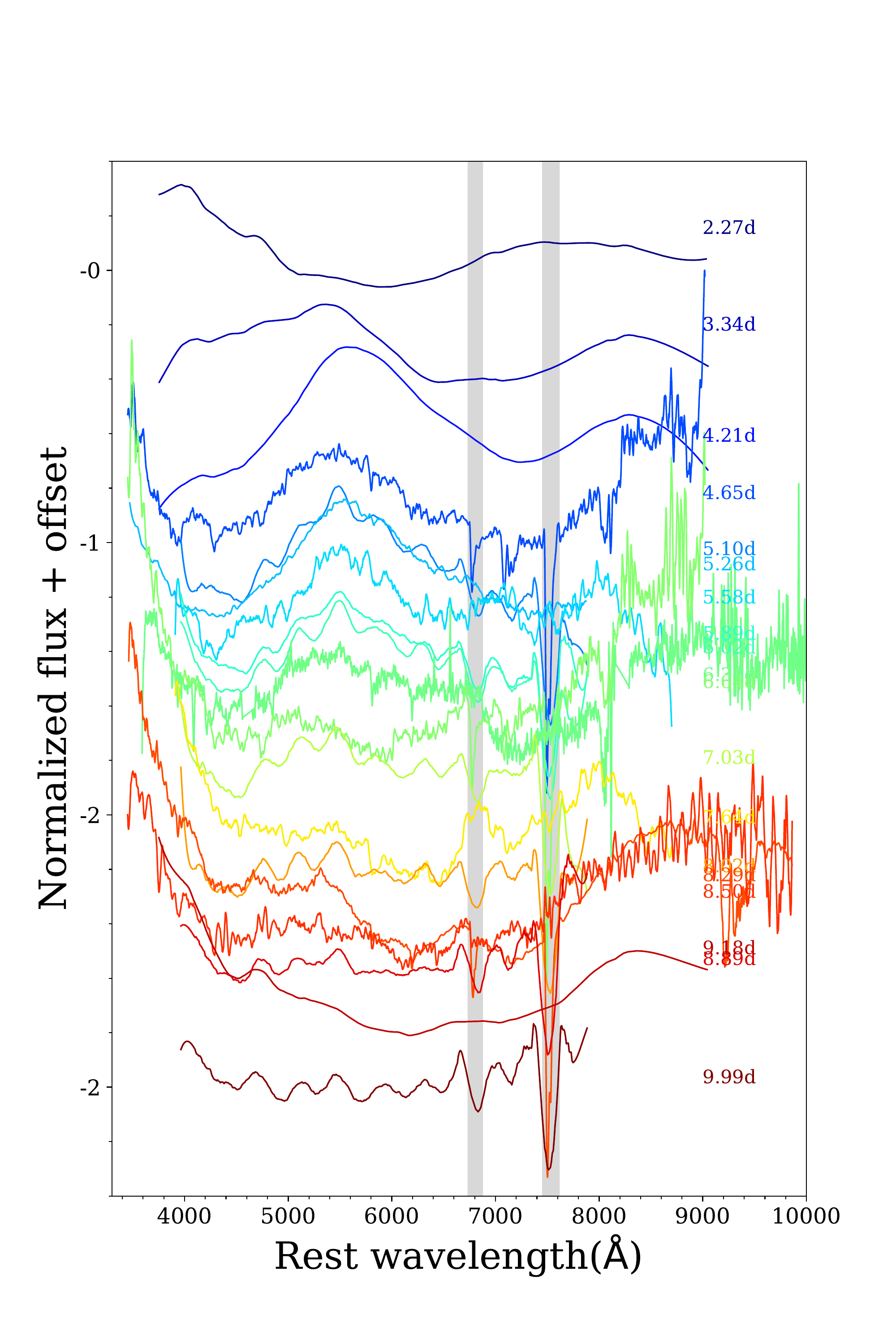}
\caption{Normalized spectra of AT2018cow in the first 10 days. The shaded areas mark the region of telluric lines. \label{fig:spectra_feat1}}
\end{figure}

The spectra of AT2018cow are characterized by featureless blue continuum in the first $\sim$10 days after discovery, and then some broad emission features emerge later, with possible contaminations from the host galaxy.
Featureless and blue spectra are common in SNe due to high photospheric temperature at early phases. Then spectral lines appear as the temperature decreases.
We create normalized spectra of AT2018cow from the observed spectra by subtracting and deviding the best fit single blackbody continuum of each spectrum.
In the first 10 days, the spectra are characterized by a wide feature near 5000\AA, as shown in Fig. \ref{fig:spectra_feat1}.
Later on, many broad emission lines emerge, overlapped with many narrow and strong emission lines.
And there is flux excess in the red end, which is probably due to dust emission in later phases.
As proposed by \cite{2019MNRAS.488.3772F}, the spectra of AT2018cow might have shown signatures of circurmstellar interaction (CSI) like SNe~Ibn and IIn.
While the typical features of CSI are narrow emission lines of H and He.
The last spectrum taken by HET shows many narrow emission lines (FWHM~$\approx$~4~\AA) which are apparently from the background host galaxy.
Although other spectra of AT2018cow do show strong and broader H$\alpha$ lines since day 8 (Fig. \ref{fig:spectra}), it is quite possible that the lines of H are from the host galaxy, not AT2018cow. The reason is that those spectra do not have such high resolution as that in HET spectrum, so the narrow lines are broadened.
To figure out whether the narrow emission lines are from the host galaxy or AT2018cow, we measured the FWHMs of H$\alpha$ line in each spectrum and compared it with other lines in the same spectrum.
The results show that the width of H$\alpha$ lines are only slightly broader than (by less than 10~\AA, within the uncertainty) other narrow lines such as [NII] and [SII], indicating that they are probably from the host galaxy.
Thus we conclude that there is no significant narrow emissions of H in AT2018cow.

\begin{figure}
\centering
\includegraphics[width=1.0\linewidth]{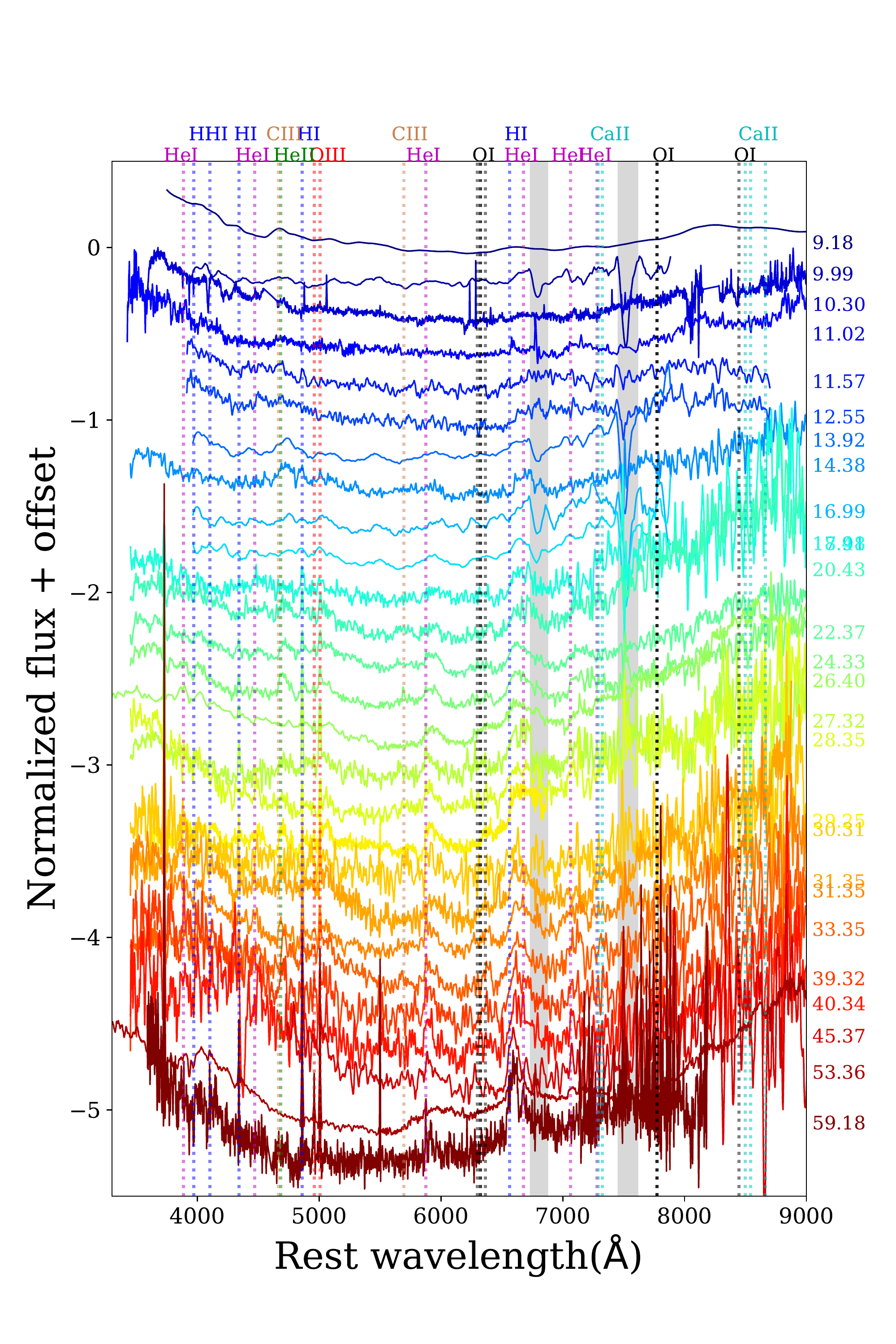}
\caption{Normalized spectra of AT2018cow after 10 days since discovery. The spectra are normalized and the narrow emission lines from the host galaxy in the ragion around H$\alpha$ in spectra of AT2018cow are manually subtracted for better view. The numbers on the right mark the time in days since discovery (MJD~58285).\label{fig:spectra_feat2}}
\end{figure}

To better look into the spectral features of AT2018cow at t>10~days, we carefully subtracted the narrow emission lines of H$\alpha$, [NII]~$\lambda$6548,6583 and [SII]~$\lambda$6730,6716 from the spectra.
For the spectra taken from $\sim$10 days to $\sim$59 days after discovery, we identify shallow and broad emission lines that can be attributed to HI, HeI, HeII, OI, OIII and CIII lines (as shown in Fig.~\ref{fig:spectra_feat2}). The OI, OIII, CIII and HeII lines dissipated after around day 45. 
The peaks of these lines are all slightly redshifted by up to 2000~km~s${^{-1}}$.
The emission lines of AT2018cow are much broader than most SNe~Ibn and IIn.
The HeI~$\lambda$5876 line has an FWHM of $\sim$300\AA\ ($v\sim15,000~\mathrm{km~s^{-1}}$) at day~14, which is one magnitude higher than that of most SNe~Ibn ( $v\sim$1000~$\mathrm{km~s^{-1}}$).
In late phases, the broad lines become narrower, with the FWHM decreasing to $\sim$3000~km~s$^{-1}$ on day 59.
Meanwhile, these broad emission lines are redshifted with velocities decreasing from $\sim$1800~$\mathrm{km\ s^{-1}}$ when they first emerge, to hundreds of km~s$^{-1}$ in late phases.
In the region of H$\alpha$, there is a broad emission line, which should be a blending of H$\alpha$ and HeI~$\lambda$6678. This line is seen getting narrower over time and splitted into two lines since t$\sim$30~day, and the peaks moves to the rest wavelength.
In addition to the long existing broad emission lines, weak and narrow (FWHM$\sim$800$-$1000~km~s$^{-1}$) HeI~$\lambda$6678 line emerged in the spectra since t$\sim$20 days. This narrow line is certain to be from AT2018cow, as it does not appear in the spectrum of the host galaxy.
To conclude, the broad emission lines of highly ionized elements (CIII, OIII) indicate that there is possible CSM interaction at very early time (t<10 days). And the appearance of narrow He emission lines in late times (t>20 days) implies the existence of another distant CSM formed around the progenitor object. 

\begin{figure}
\centering
\includegraphics[width=1.0\linewidth]{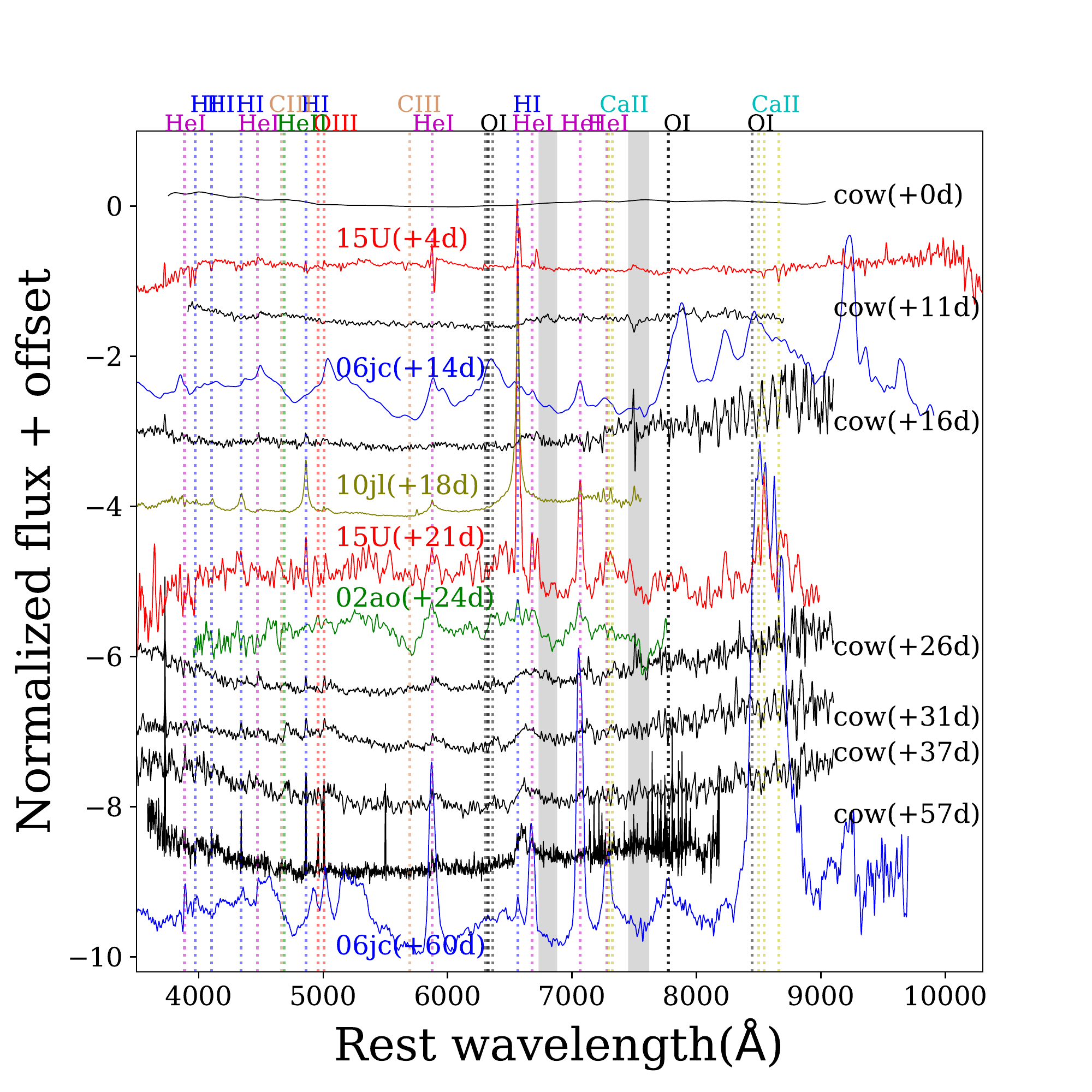}
\caption{Normalized spectra of AT2018cow compared with other supernovae. All spectra are normalized by the fitted blackbody continum. The numbers in the brackets are the phases relative to $V$-band maxmum, which for AT2018cow we adopt MJD~58287. The narrow H$\alpha$ lines in spectra of AT2018cow are manully subtracted for better view.\label{fig:spectra_cmp}}
\end{figure}

It is natural to think of an interacting SN picture for AT2018cow.
\cite{2019MNRAS.488.3772F} found the similarity between AT2018cow and some SNe~Ibn and SNe~IIn.
Here we argue that although AT2018cow show signitures of interaction similar to SNe~Ibn and SNe~IIn, its spectral evolution is quite different from that of SNe~Ibn and IIn.
In Fig.~\ref{fig:spectra_cmp} we show the spectral evolution of AT2018cow compared with some well observed SNe~Ibn, SN~2006jc \citep{2007Natur.447..829P, 2008ApJ...680..568S}, SN~2015U \citep{2015MNRAS.454.4293P, 2016MNRAS.461.3057S} and SN~2002ao \citep{2008MNRAS.389..113P}, and a typical SN~IIn 2010jl \citep{2012AJ....143...17S, 2012AJ....144..131Z}.
From Fig.~\ref{fig:spectra_cmp} we can see the diversity of SNe~Ibn. AT2018cow seems to have different spectral features from any other interacting SNe, as it has weaker lines at all phases.
At earlier phases AT2018cow is characterized by blue featureless continum like that seen in some CCSNe as a result of hight temperature, i.~e. SN~2015U from our comparison sample. While SN~2015U shows a narrow P-cygni absorption feature, indicating the recombination of He in the CSM \citep{2016MNRAS.461.3057S}.
Note that the emission lines of AT2018cow emerged at later phases, and are much weaker compared to SN~2006jc and SN~2002ao.
Moreover, the Ca lines are very strong in SN~2006jc and SN~2015U, but are weak in AT2018cow.
At late times, AT2018cow show similarities to SN~2002ao, both being dominated by broad lines. While P-cygni absorption of HeI lines are present in SN~2002ao, and the lines are stronger. The line velocity of HeI at t$\sim+$24d is $\sim$8500~km~s$^{-1}$ for SN~2002ao, slightly higher than AT2018cow ($\sim$7000~km~s$^{-1}$).
SN~2002ao is claimed as 06jc-like, which are proposed as Wolf-Rayet (WR) stars exploded in a He-rich CSM \citep{2008MNRAS.389..113P}.
Although SNe~Ibn and IIn are distinguished by the strength of the H emission lines, there are some transitional objects which show roughly equal strength of H and He emission lines, for example SN~2005la \citep{2008MNRAS.389..131P} and SN~2011hw \citep{2012MNRAS.426.1905S, 2015MNRAS.449.1921P}.

In most interacting supernovae, like SNe~IIn and Ibn, the emission lines have velocities in range of tens to a few thousand km~s$^{-1}$, depending on the wind velocities of the progenitor stars. The wind velocities are related to the type of the progenitor stars. At same metallicities, stars with larger initial masses are expected to have stronger stellar winds therefore higher wind velocities when they evolve to the end of life (see \citealp{2014ARA&A..52..487S} and references therein).
Some of the objects show intermediate width emission lines (1000~km~s$^{-1}$~<~$v$~<~4000~km~s$^{-1}$), like SN~2006jc.
In spectra of SN~2006jc, the bluer He lines show narrow P-Cygni profiles, while the redder He lines show an intermediate width emission component (FWHM$\approx$~3000~km~s$^{-1}$) \citep{2007ApJ...657L.105F}.
The broad emission features in AT2018cow are apparently different from the spectral features in ordinary SNe~II. The lack of absorption features implies that AT2018cow is possibly more similar to SNe~IIn/Ibn, rather than SNe~IIP/IIL.
While the velocities of the broad emission component in AT2018cow ($v\sim$10,000~km~s$^{-1}$) are much higher than normal SNe~Ibn/IIn.
The lack of narrow emission lines in AT2018cow and relatively weak lines make it unique among interacting SNe. While this is not an argument against the interacting SN origin of AT2018cow, because spectral diversity is seen in other SNe~Ibn and SNe~IIn \citep[e.g., ][]{2017ApJ...836..158H}. Absence of narrow lines might be resulted from a closely located CSM which was immediately swept up by the shock within a short time period. 
\deleted{This may be a result of the CSM being accelerated by the supernova shock. At later times the CSM slows down so the lines become narrower.}

\section{Host galaxy environment}\label{sec:host_env}
\begin{figure}
\centering
	\begin{minipage}{1.0\linewidth}
    \includegraphics[width=1.\linewidth]{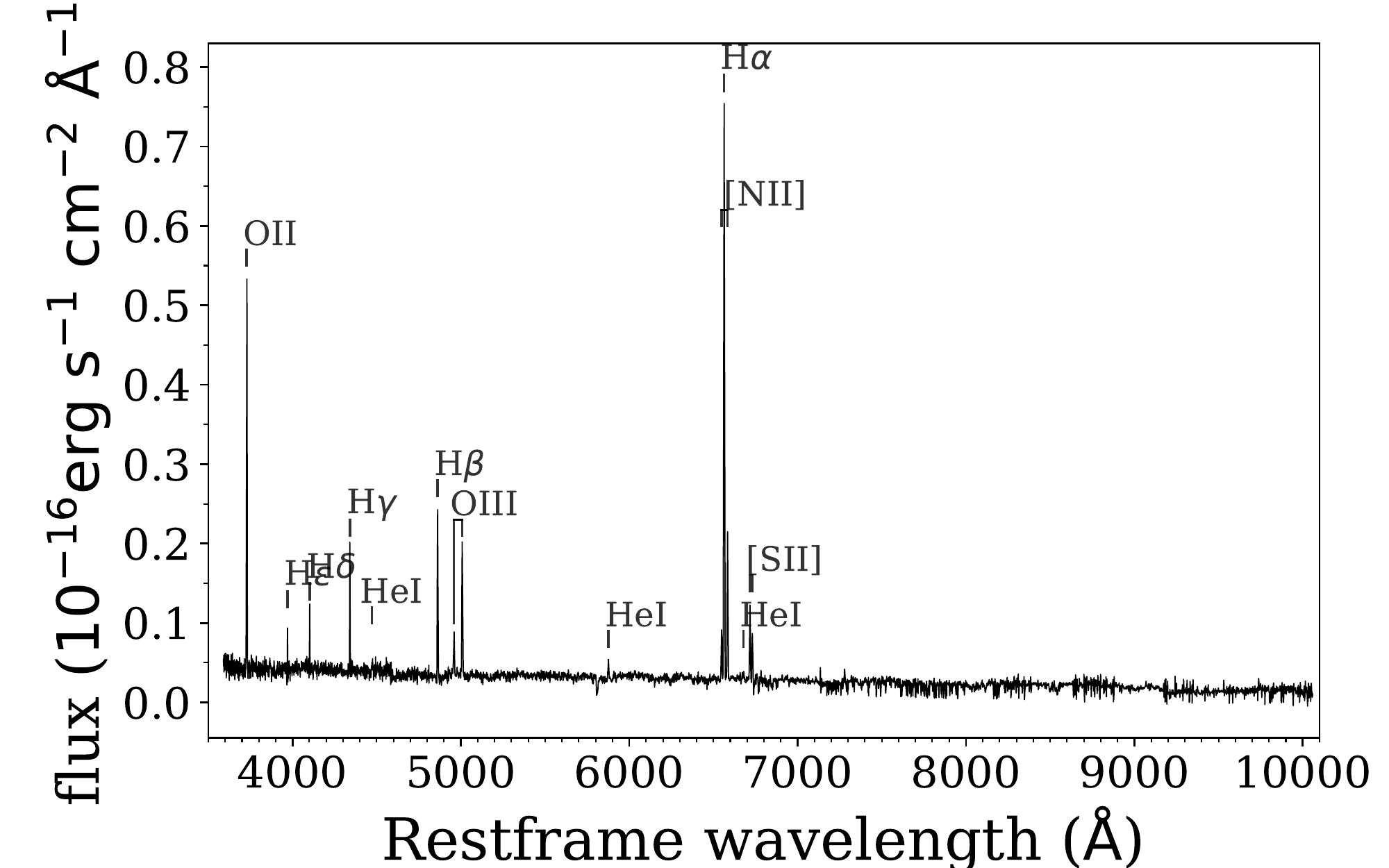}
    \end{minipage}

    \begin{minipage}{1.0\linewidth}
	\includegraphics[width=1.0\linewidth]{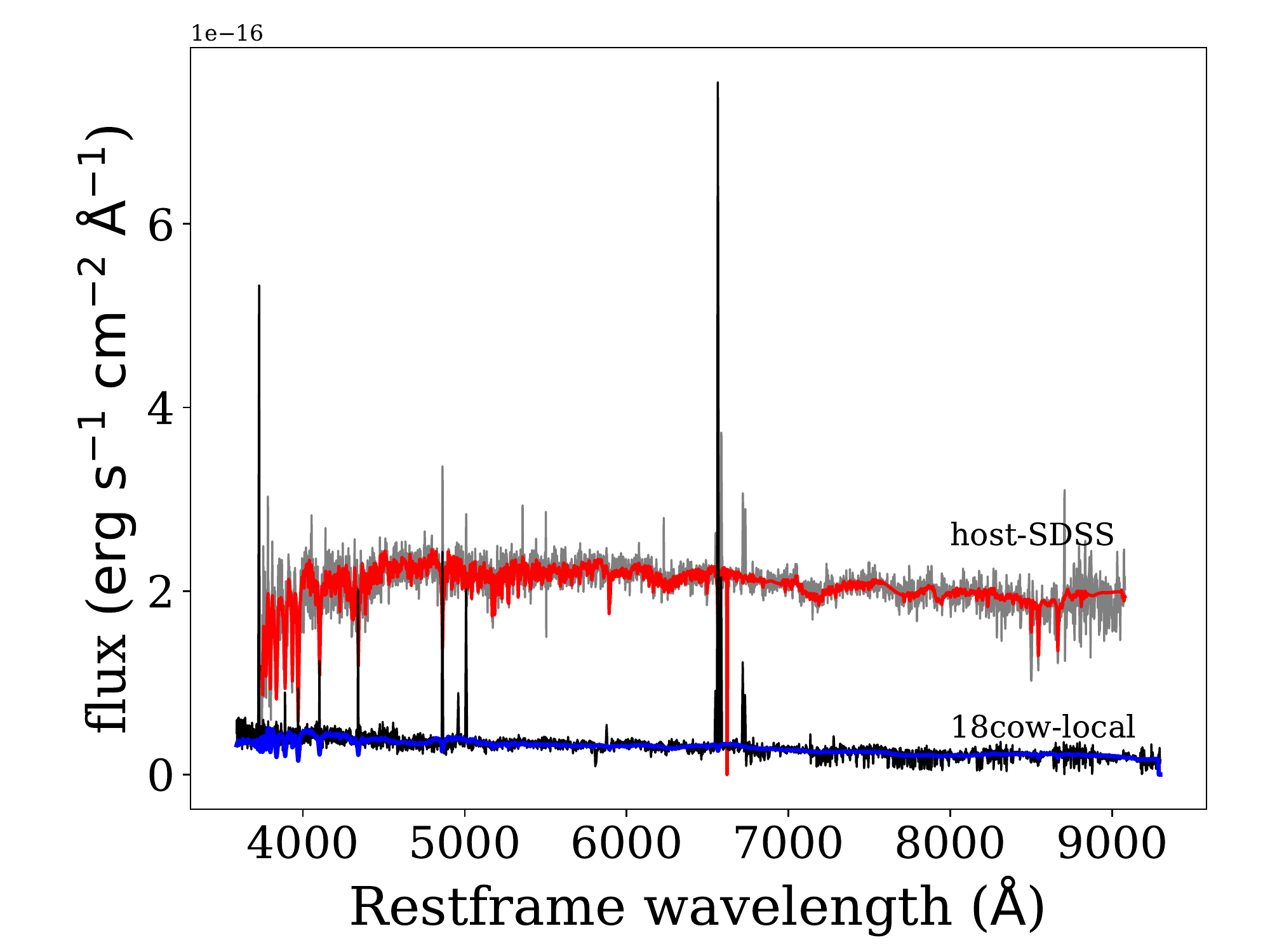}
	\end{minipage}
\caption{Upper panel: The spectrum taken at the lcation of AT2018cow. Lower panel: Firefly fitting of the host spectra at the site of the object and galaxy center. The over-plotted colored lines are the best fit models of Firefly. \label{fig:host_spec}}
\end{figure}

We notice that the spectra of AT2018cow are almost featureless at early phases (t<10~d).
Later on the spectra are some broader features overlapped with many narrow emission lines which are most probably due to the emission from the background galaxy.
We obtained a spectrum of the host at the location of AT2018cow with the 9.2-m HET on Sep. 17, 2019 (corresponding to $\sim$460 days after discovery), as shown in Fig. \ref{fig:host_spec}.

The spectrum is characterized by that of a typical HII region, which implies that this region is currently at gas phase and star-forming. One can see strong emission lines of H, He, N, S and O, and a NaI~D absorption line at the rest wavelength of the Milky Way.
With this spectrum, we are able to measure the intensities of the emission lines and then derive the properties of the local environment.
Following \cite{2017MNRAS.465.1384C}, we get a local metallicity of 12+log(O/H)$\sim$8.65$\pm$0.07, which is solar-like and among the range of other SNe~Ibn \citep{2016MNRAS.456..853P}.
The star formation rate (SFR) can be derived from the luminosity of the H$\alpha$ emission line, for which we measured as $L$(H$\alpha$)$\approx1.82\times10^{39}$~erg~s$^{-1}$.
This is consistent with the result from \cite{2020MNRAS.495..992L} $L$(H$\alpha$)$\approx1.35\times10^{39}$~erg~s$^{-1}$ at the site of AT2018cow, considering that we applied a larger distance.
Using the conversion factor given in \cite{2001ApJ...558...72S}, we get SFR(H$\alpha$)$_{local}\approx 0.015$~\msun~yr$^{-1}$.
We also examine the [OII]$\lambda$3727 line, and get $L$([OII])$\approx 8.76\times 10^{38}$~erg~s$^{-1}$. With the relation given in \cite{1998ARA&A..36..189K}, we get SFR([OII])$_{local}\approx 0.012$~\msun~yr$^{-1}$, which is consistent with that from H$\alpha$ line.
To get more information of the local environment of AT2018cow, we use Firefly \citep{2017MNRAS.472.4297W} to fit the spectrum with stellar population models.
The input models are two M11 libs: MILES and STELIB \citep{2011MNRAS.418.2785M}, and initial mass function `Kroupa' \citep{2001MNRAS.322..231K} is adopted in the fit.
Fig.~\ref{fig:host_spec} shows the best fit specta, from which we get a stellar mass of $M_{\star}\sim5\times10^6$~\msun.
Combining the above SFR and stellar mass infomation, we can get a local specific star formation rate (sSFR) as log(sSFR)$_{local}\sim-$8.5~(yr$^{-1}$). 

\deleted{Using Equation 4 from \cite{2003ApJ...591..827P}, which makes use of the flux ratio of the emission lines in galaxies, we obtained a star formation rate (SFR) of 0.821$\pm$0.215 \msun\ $\mathrm{yr^{-1}}$.
While a much lower SFR, i.e., $\sim$0.22~\msun\ $\mathrm{yr^{-1}}$, was derived by \cite{2019MNRAS.484.1031P}, likely due to that different methods were used in deriving the SFR. \cite{2019MNRAS.484.1031P} fits the SED of the host galaxy with stellar population models.}

\deleted{To get more information of the host environment of AT2018cow, we use Firefly \citep{2017MNRAS.472.4297W} to fit the spectrum with stellar population models.
The input models are four M11 libs: MILES, STELIB, ELODIE, and MARCS \citep{2011MNRAS.418.2785M}, and initial mass function `Kroupa' \citep{2001MNRAS.322..231K} is adopted in the fit.
Fig.~\ref{fig:host_spec} shows the best fit specta, from which we get a stellar mass of $M_{\star}\sim(3.15\pm1.69)\times10^7$~\msun\ and an age of $0.78\pm0.20$~Gyr.}

The Sloan Digital Sky Survey \citep[SDSS; ][]{2018ApJS..235...42A} has taken one spectrum at the center of the host galaxy of AT2018cow on MJD~53566.
As the HET spectrum we obtained only provides the local information, we also use the SDSS spectrum to measure the above corresponding parameters for the whole galaxy.
The resulting metallicity is the same as that measured from the HET spectrum spotted at the site of AT2018cow, while the SFR is measured as SFR(H$\alpha)_{center}\approx0.008$~\msun~$\mathrm{yr^{-1}}$ if we do not consider any host extinction.
The Firefly fit shows that the stellar mass of the nucleus is $M_{\star}\sim2.6\times10^8$~\msun.
We caution that the SDSS spectrum only includes the flux from the galaxy center, thus the SFR is expected to be lower. For the whole galaxy, we refer to the results from other studies. \cite{2019MNRAS.484.1031P} and \cite{2020MNRAS.495..992L} found stellar mass and SFR in good agreement with each other, although they adopted different distances. At $D_L$=63~Mpc, stellar masses in these two studies become $M_{\star}\approx1.56\times10^9$\msun\ and $1.85\times10^9$\msun, respectively. And the SFR from \cite{2020MNRAS.495..992L} becomes 0.20~\msun~yr$^{-1}$. In the following discussion, we adopt an average of these results, i.e. $M_{\star}\approx1.70\times10^9$\msun, SFR$\approx$0.21~\msun~yr$^{-1}$, and log(sSFR)$\approx-$9.88~(yr$^{-1}$).

\begin{figure}[htb]
\includegraphics[width=1.2\linewidth]{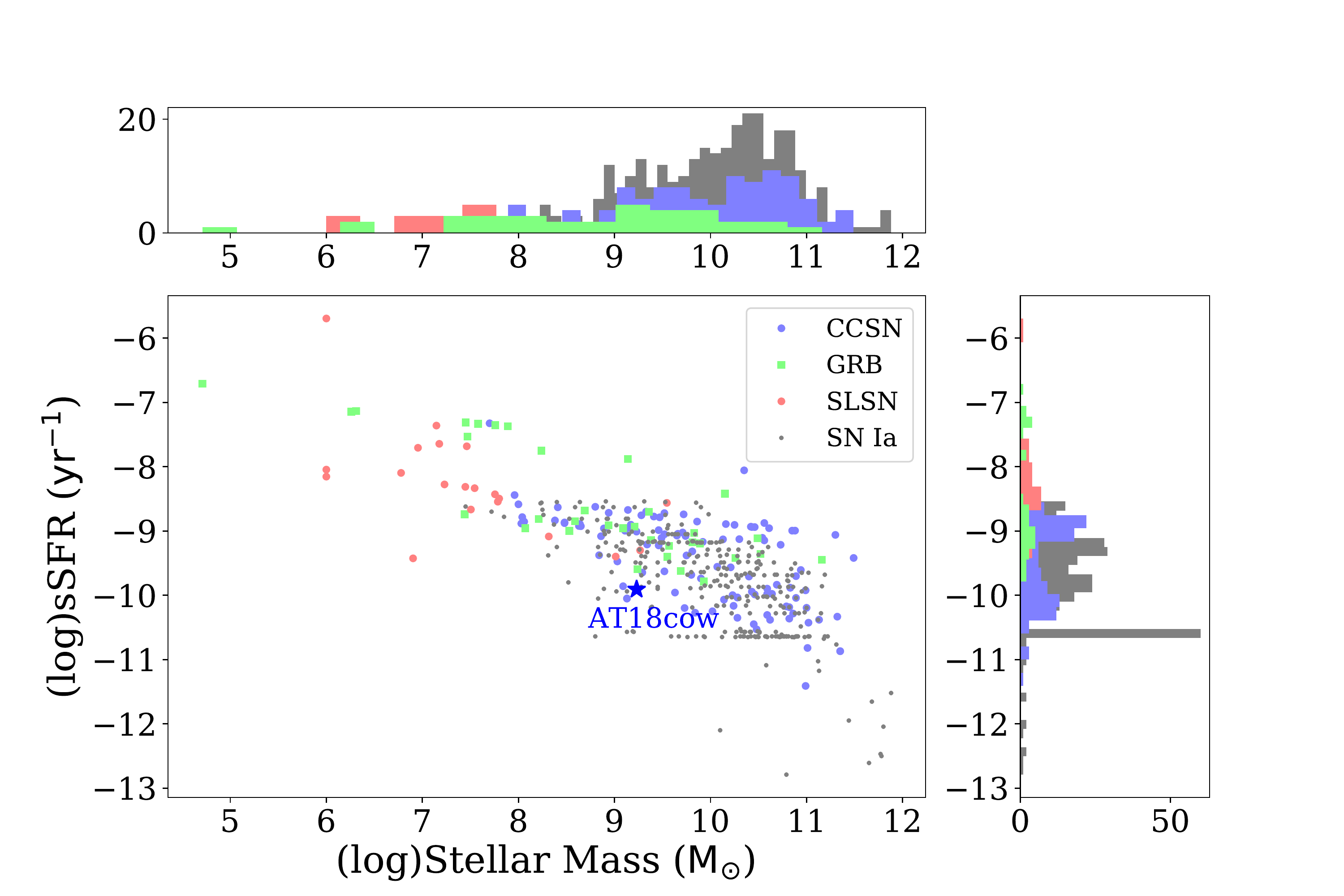}
\caption{The host-galaxy parameters of AT2018cow compared with other types of transients. \label{fig:host_env}}
\end{figure}

The host environment may provide a clue to the physical origin of AT2018cow. We compare the host environment parameters with other well studied transients, including type Ia supernovae \citep[SNe~Ia, ][]{2012ApJ...755...61S, 2014A&A...572A..38G}, core-collapse supernovae \citep[CCSNe, ][]{2010MNRAS.405...57S, 2014A&A...572A..38G}, superluminous supernovae \citep[SLSNe, ][]{2016MNRAS.458...84A}, and gamma-ray bursts \citep[GRBs, ][]{2010MNRAS.405...57S}.
\deleted{As shown in Fig.~\ref{fig:host_env}, the host of AT2018cow is located in the range of CCSNe and GRBs, but it is located far from the center of the SLSNe and SNe~Ia groups. The host galaxies of SLSNe have both lower stellar mass and SFR, indicating that there might be a distinction between CCSNe and SLSNe. The host galaxy of AT2018cow is located among CCSNe, SNe Ia and GRBs, implying that AT2018cow might arise from massive stars.}
As shown in Fig.~\ref{fig:host_env}, the host galaxy of AT2018cow is located among SNe Ia, CCSNe and GRBs, but away from the SLSNe group. The host galaxy of AT2018cow has stellar mass close to the median of GRBs, but at the lower end of the SNe group, except for SLSNe. We can not say for sure which group it should belong to, and it is likely that AT2018cow is distinct from SLSNe, although AT2018cow has a peak luminosity comparable to them. Meanwhile, the local high SFR of AT2018cow may imply that AT2018cow is probably originated from a massive star.

\section{Modeling the rapid evolving light curves}\label{sec:lc_model}
The physical interpretation of AT2018cow is still in debate, although there are already several papers trying to uncover its physical origin.
The radioactive decay of \Ni is a well known energy source for supernovae \citep{1982ApJ...253..785A}.
The bolometric light curve of AT2018cow can not be powered by pure \Ni, as the peak luminosity would require an ejected \Ni mass of $\sim$~6~\msun\ but a low ejecta mass <~1~\msun.
In the above analysis, we find high resemblance of the light curves of AT2018cow to that of SNe~Ibn, and signatures of CSI are found in the spectra, so we try to fit the light curves of AT2018cow using the CSI model.
The fast-declining and luminous bolometric light curve of SN~2006jc has been successfully modeled by CSI models \citep[e.g., ][]{2009MNRAS.400..866C, 2008ApJ...687.1208T}.
The rapid declining light curves can be related to the early shock-cooling from the progenitor envelope. Since the progenitor has lost most of its hydrogen envelope, the shock-cooling should be weak and short for the core-collapse of a massive star.
Another reasonable interpretation is the interaction of the supernova ejecta with the surrounding circumstellar medium (CSM). This can be supported by the emission lines in the spectra (see Sec. \ref{subsec:spectral_evolution}).

\begin{figure}
\centering
\includegraphics[width=1.0\linewidth]{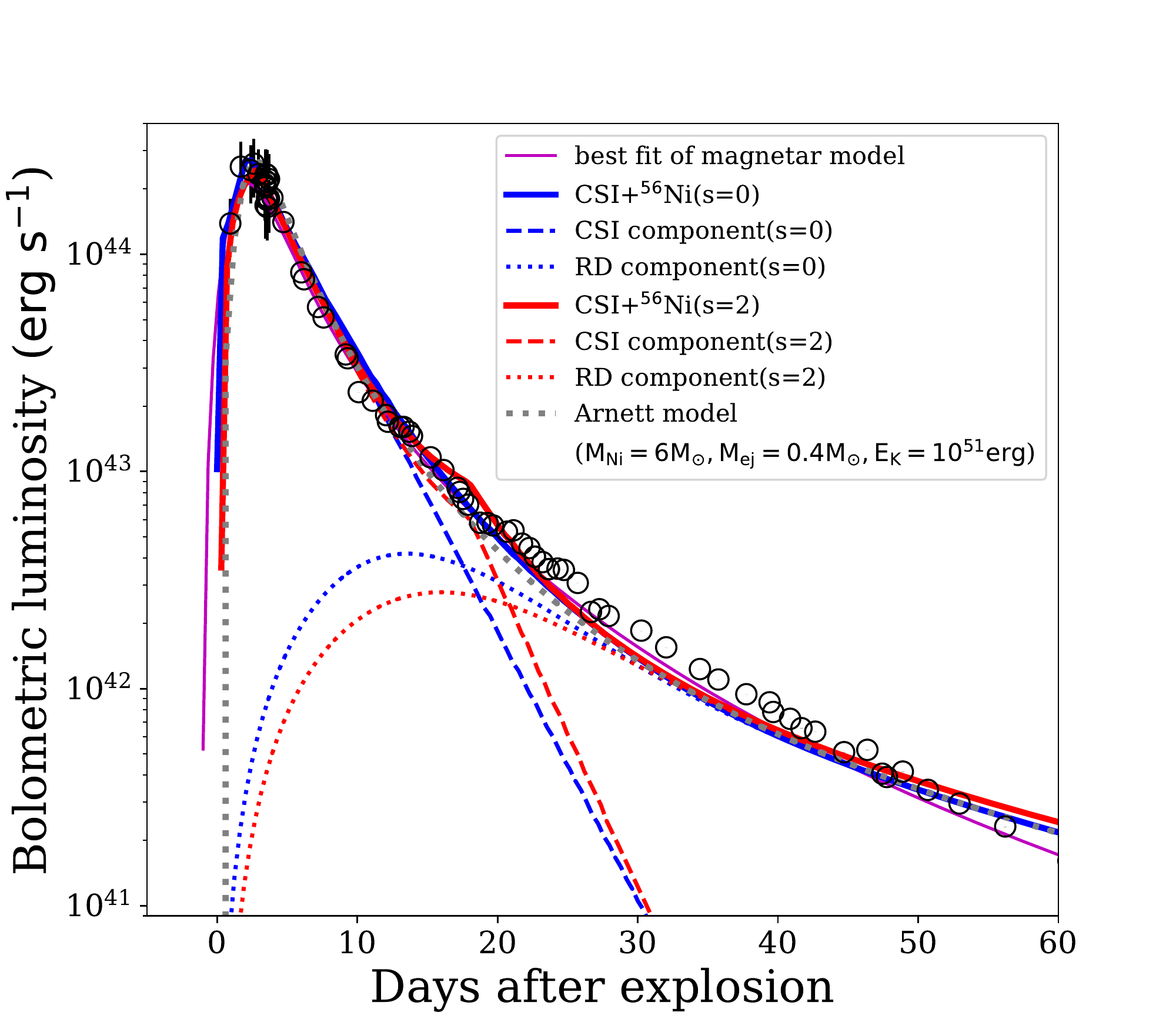}
\caption{Bolometric light curve of AT2018cow. Best-fit models of CSI+RD with s=0, 2 are plotted in blue and red solid lines, respectively. The two components of the models are shown by dashed (CSI) and dotted (RD) lines, respectively. The best-fit magnetar model is plotted as a magenta solid line. A pure RD model is also shown in a grey dotted line as a reference. \label{fig:lc_bol}}
\end{figure}

We construct the bolometric light curves by integrating the UV and optical flux (the UV data are taken from \citealp{2019MNRAS.484.1031P}), and then apply a model in which CSI is dominating the early time light curve.
In order to constrain the fitting better, especially to obtain data before peak, we estimate the pre-peak bolometric luminosities based on the following assumptions: 1). the SED of AT2018cow is a blackbody; 2). the photometric temperature evolves as a power law $T\propto (t-t_0)^{-0.5}$, as we derived from the early temperature evolution. And then the bolometric luminosities before MJD~58288.44 are estimated using the single band photometry data.
We adopt a hybrid model which includes \Ni powering and the interaction of the SN ejecta with a dense CSM with density profile as a power-law, i.~e. $\rho_{\mathrm{CSM}}\propto r^{-s}$, where the typical value of s is 2 and 0 \citep[e.g., ][]{2012ApJ...746..121C, 2013ApJ...773...76C, 2019MNRAS.489.1110W}. In our model, the density distribution of the ejecta is uniform in the inner region ($\delta$=0), and follows a power-law ($\rho\propto r^{-12}$) in the outer region. The early fast rising light curve of AT2018cow is mainly powered by CSI, while the slower declining tail is dominated by radioactive decay (RD) of \Ni.
We first consider the case of s=2, which corresponds to a steady-wind CSM.
The best-fit light curve is shown in Fig.~\ref{fig:lc_bol}, and the fitted parameters are presented in Tab.~\ref{tab:fit_parms}.
As shown in Fig.~\ref{fig:lc_bol}, our CSI+RD(s=2) model can fit the observations quite well. The mass loss rate of the progenitor star can be estimated as $0.1(v_\mathrm{CSM}/100$~km~s$^{-1}$)~\msun~yr$^{-1}$, $\sim$~1~\msun~yr$^{-1}$ with $v_{\mathrm{CSM}}\sim$ 1000~km~s$^{-1}$. \cite{2019ApJ...872...18M} also reach similar conclusion by analysing the optical and X-ray data. Such a mass loss rate is much higher than that found from the radio observations of AT2018cow ($\dot{M}\sim10^{-4}-10^{-3}$~\msun~yr$^{-1}$) (\citealp{2019ApJ...871...73H} had a similar conclusion). If we set limit on the mass loss rate, the model can hardly fit the observations. Thus, we claim that the early bright and fast evolving light curve of AT2018cow can not be produced by CSI with a steady stellar wind.

We then try the other case where s=0, i.~e. the density of the CSM is a constant.
The fitting result is shown in Fig.~\ref{fig:lc_bol}, and the fitted parameters are presented in Tab.~\ref{tab:fit_parms}.
As shown in Fig.~\ref{fig:lc_bol}, with $M_{\mathrm{ej}}\approx3.16$~\msun, $M_{\mathrm{CSM}}\approx0.04$~\msun, $M_{\mathrm{Ni}}\approx0.23$~\msun, the CSM+\Ni(s=0) model can also provide a plausible fit for the observed bolometric light curve.
The inferred inner radius of the CSM gives a constraint on the radius of the progenitor star $R<3~\mathrm{R_\odot}$, which is consistent with the typical size of WR stars.
The CSM shell extends outwards to a radius of $8.70\times10^{13}$~cm ($\approx1200~R_{\odot}$), implying that the CSM was formed shortly prior to the explosion.
\deleted{The WR stars cannot generate such large mass loss, while binary interaction is a possible mechanism to account for such a large amount of mass ejection per year, i.e., $10^{-3}-10^{-1}$~\msun~yr$^{-1}$ (see \citealp{2019MNRAS.488.3783B} and references therein). The inner radius of the CSM implies that the CSM is formed shortly prior to the explosion.}
Such a CSM shell can be produced by an episodic mass ejection from the progenitor star, like a luminous blue variable (LBV) or from a common-envelope episode of a binary system.
Combining the mass and velocity of the ejecta, the kinetic energy of the ejecta can be estimated as 6.6$\times10^{51}$~erg, several times higher than that of the ordinary SNe~Ibc and rather similar to the broad lined SNe~Ic (SN~Ic-BL) \citep{2016MNRAS.457..328L}, which are found to be possibly associated with long gamma ray bursts (e.g. SN~1998bw \citep{1998Natur.395..672I, 2001ApJ...550..991N}). The high velocity of the ejecta might be connected to a relativistic jet.

\begin{deluxetable}{lcc}
\hspace{-1.0cm}
\tablecolumns{3}
\tablewidth{0pc}
\tablecaption{Parameters of the best fit CSM+$^{56}$Ni models for AT2018cow.\label{tab:fit_parms}}
\tablehead{\colhead{...} &\colhead{s=0} &\colhead{s=2}}
\startdata
$t_0$                         & MJD 58284.5  & MJD 58284.7   \\
$M_\mathrm{ej}(M_\odot)$      &3.16    &1.69\\
$x_0$\tablenotemark{a}        &0.63   &0.76\\
$v_\mathrm{ej}$(km~s$^{-1}$)  &26000   &13600 \\
$M_\mathrm{Ni}(M_\odot)$      &0.23    &0.14 \\
$M_\mathrm{CSM}(M_\odot)$     &0.04    &0.12 \\
$r_\mathrm{CSM,in}$(cm)\tablenotemark{b}  &2.11$\times10^{11}$   & 1.03$\times10^{13}$ \\
$r_\mathrm{CSM,out}$(cm)\tablenotemark{c} &8.70$\times10^{13}$   & 3.16$\times10^{14}$  \\
$\rho_\mathrm{CSM,in}$(g~cm$^{-3}$)\tablenotemark{d}  & $2.9\times10^{-11}$   & $5.6\times10^{-10}$  \\
$\epsilon$\tablenotemark{e}  & 0.22 & 0.65 \\
$\kappa$  & 0.14  & 0.15  \\
$\kappa_{\gamma}$  &0.015  &0.014
\enddata
\tablenotetext{a}{The dimensionless radius of the division between inner and outer region of ejecta.}
\tablenotetext{b}{The inner radius of CSM.}
\tablenotetext{c}{The outer radius of CSM.}
\tablenotetext{d}{The CSM density at $r_\mathrm{CSM,in}$.}
\tablenotetext{e}{The radiation efficiency.}
\end{deluxetable}

Alternatively, the bolometric luminosity and effective temperature evolutions can be explained by a magnetar-powered model \citep{2017ApJ...850...55N}. Assuming that the opacities of ejecta are $\kappa=0.2~\mathrm{cm^2~g^{-1}}$ for the optical photon and $\kappa_\mathrm{mag}=0.013~\mathrm{cm^2~g^{-1}}$ for the magnetar wind, respectively, the best-fit parameters for this model are $t_0=58283.4$, $M_\mathrm{ej}=0.1$ \msun, $v_\mathrm{ej}=2.72\times10^4$ km~s$^{-1}$, $P=4.5$ ms and $B=1.11\times10^{14}$ G, where $P$ and $B$ are the initial spin period and magnetic field strength of the nascent magnetar, respectively. We caution that the best-fit $M_{\mathrm{ej}}$ = 0.1~\msun\ is the lower limit of the magnetar model in our fitting program. If no limit is given, the fitting tends to find a significantly lower value to fit the narrow light curve better. This may imply that the magnetar-powered model requires a rather low ejecta mass for AT2018cow.

\section{Discussion: Progenitor Properties}\label{sec:discussion}
In the previous section, we made analysis of the bolometric light curve of AT2018cow based on an assumption that it is a supernova origin.
While we do not rule out other possibilities, especially the TDE origin.
A main problem of the supernova origin for AT2018cow is that the process of an expanding photosphere is missing. In early phases, the photospheric velocity may be very high ($\sim$0.1c) for AT2018cow in early phases. The photospheric radius keeps decreasing since very early time.
This can be a clue for the interpretation as a TDE for AT2018cow. Although both \cite{2020MNRAS.495..992L} and \cite{2019ApJ...872...18M} find no evidence of the connection between the site of AT2018cow and an IMBH.
Nevertheless, one can notice that the measurements of photospheric radius start after the peak, probably suggesting that the expanding phase is not observed.

The magnetar-powered model can make a good fit to the bolometric light curve.
\deleted{It can also fit the temperature evolution when we assume the final plateau temperature $T_\mathrm{f}$ to be $\sim$ 13,000~K.
While $T_\mathrm{f}$ is found to be in range of $\sim$4000-9000~K for SLSNe \citep{2017ApJ...850...55N}. The $T_\mathrm{f}$ for AT2018cow is far higher than most cases seen in SNe.}
The best-fit $B$ and $P$ of the central engine lies in the range of SLSNe (\citealp{2020ApJ...903L..24L} and references therein).
Distinction between At2018cow and SLSNe is the evolution timescale, which is related to the ejecta mass.
\cite{2017ApJ...850...55N} found $M_{\mathrm{ej}}\geqslant2.2$\msun\ with an average of 4.8\msun.
Besides, the low ejected mass ($M_{\mathrm{ej}}\sim0.1$\msun) required by the magnetar model for AT2018cow is not likely favorable for a massive star, except for some really extreme cases.
Some studies find that massive stars can be ultra-stripped by binary interaction with a compact neutron star \citep{2015MNRAS.451.2123T}. But in these cases, little H or He remains in the progenitor system, which is not consistent with the observed spectral features of AT2018cow. 
Thus, we disfavor the magnetar model for AT2018cow.

Our CSI+RD(s=0) model makes a plausible fit to the bolometric light curve of AT2018cow.
With $R<3R_{\odot}$, the progenitor star is most likely to be a compact WR star. The ejected mass ($M_{\mathrm{ej}}\approx$3\msun) is lower than that predicted by single stellar evolution models \citep[e.g., ][]{2012A&A...542A..29G} but around the mean value of SNe~Ibc \citep{2016MNRAS.457..328L}. This might be a result of binary interaction or episodic eruptive mass loss during the lifetime of massive stars. It is hard to derive the mass of the progenitor star simply from the ejecta mass, since the mass loss mechanisms of massive stars can be complicated and ambiguous.

In the case of s=0, the CSM can be dense shells formed by strong stellar winds of WR stars or an eruptive of LBV stars \citep{1989ApJ...344..332C, 2011MNRAS.412.1639D}. According to our fitting result, with wind velocity of 100~km~s$^{-1}$, the eruption started several months before core-collapse, and was possibly still on when exploding. The average mass loss rate is $\sim$0.15\msun~yr$^{-1}$, or even higher if the wind velocity is higher, lying well in the range of LBV eruptions \citep{2014ARA&A..52..487S}. Such mass loss behaviour can be found in some SNe~IIn and SNe~Ibn\citep{2007ApJ...656..372G, 2015A&A...580A.131T, 2015MNRAS.449.1921P, 2012ApJ...744...10K, 2014MNRAS.439.2917M}. Under this scenario, the progenitor of AT2018cow might be a massive star which is during eruptive state. However, with $R<3R_{\odot}$, the progenitor star is most likely to be H-poor or even He-poor, so is the CSM. While there is possibility that H and He are mixed into inner shells so that the progenitor can keep some H/He at core-collapse. 

Meanwhile, binary interaction might dominate the evolution of massive stars, which are thought to be the progenitors of stripped envelope supernovae (SESNe). Mass loss can be quite efficient in binaries \citep{2017PASA...34...58E}. SN~2006jc is a representative of interacting SNe originated from binary massive stars \citep{2016ApJ...833..128M, 2020MNRAS.491.6000S}. In the binary scenario, the progenitors can be less massive stars, and the companion stars evolve slower so that they can keep their H/He envelopes. A common-envelope episode of a binary system can also form this dense CSM shell.
The detection of H and He lines in the spectra of AT2018cow indicates that the CSM is not H-free. So it is quite possible that the CSM is from the companion star rather than the progenitor itself. The slightly redshifted peaks of the emission lines in the spectra of AT2018cow suggest asymmetry of the CSM, in favor of the common-envelope picture. 
\deleted{There remains problems on the properties of the progenitor and its mass loss mechanism.
According to stellar evolution theory, the CSM of $\sim$0.7\msun\ can be produced by pulsational ejection of very massive stars, or stripped from the star by interaction with a companion star.
The mass loss rate we derived is in the range of WR stars, and similar to that of the progenitors of SNe~IIn \citep{2012ApJ...744...10K, 2014MNRAS.439.2917M}.
With a mass loss rate $\dot{M}\sim10^{-2}~\mathrm{M_{\odot}~yr^{-1}}$, it takes $\sim$~70~years to lose 0.7~\msun\ of the stellar mass.
Another possibility is that the progenitor is a luminous blue variable (LBV) star, so that it underwent occasional giant eruptive events before explosion (e.g., \citealp{2007ApJ...657L.105F}).
Alternatively, the low ejected mass ($\sim$3.5\msun) favors that the progenitor might not be a single very massive star but a less massive star in a binary system.
The narrow and weak lines of He~I which emerged at t$\sim$20~days after discovery imply that there is another CSM in the outer region, which was formed prior to the inner denser shell. This CSM might be wind-like. The interaction of this CSM and the ejecta can be the source of the radio emission observed by \cite{2019ApJ...872...18M} and \cite{2019MNRAS.484.1031P}.}

The progenitor star could be a very massive star which have experienced violent mass loss due to pulsational ejection.
Recently \cite{2020ApJ...903...66L} has proposed a scenario based on a pulsational pair-instability supernova (PPISN) model, concluding that the rapidly evolving light curve of AT2018cow can be explained by a 42~\msun\ He star exploding in a dense He-rich CSM ($M_{\mathrm{CSM}}\sim0.5$~\msun).
\deleted{The large amount of CSM derived by our CSI+RD model ($M_{\mathrm{CSM}}\sim0.7$~\msun) is close to their value.}
The proposed model can fit the bolometric light curve well (at t<30 days). However, the presence of H lines in the spectra of AT2018cow is inconsistent with the assumption that both the ejecta and CSM formed around AT2018cow should be H-poor. \cite{2020ApJ...903...66L} tested different compositions of the CSM and found that the amount of H in the CSM only has slight effect on the bolometric light curve.
Our fitting result is in agreement with \cite{2020ApJ...903...66L} in terms of the density and size of the CSM, but we find much lower CSM mass.
\deleted{The difference between our model and \cite{2020ApJ...903...66L} model lies on the mass of ejected \Ni and its distribution, the ejected total mass.}
We do not assume any Ni-mixing, while \cite{2020ApJ...903...66L} assumes that Ni is fully mixed into the outer layers of the ejecta.
Nevertheless, both models may be plausible. Our model can correspond to a massive progenitor in a binary system, while \cite{2020ApJ...903...66L} requires a very massive star, whose zero-age-main-sequence (ZAMS) mass is 80~\msun.
It is worth noting that \cite{2019ApJ...887...72L} claims that a massive He core can only be formed under low metallicity ($Z\leq0.5Z_{\odot}$), which is inconsistent with our measurement of a solar-like metallicity environment for the progenitor of AT2018cow (see Sec. \ref{sec:host_env}). This may imply that the progenitor of AT2018cow did not undergo PPI.

The fast evolving light curves of AT2018cow may be related to a very low ejecta mass, which is consistent with electron-capture supernovae (ECSNe) \citep{1984ApJ...277..791N, 1987ApJ...322..206N, 1991ApJ...367L..19N, 2014A&A...569A..57M}.
Stars with ZAMS mass of $\sim$8-12 \msun\ form degenerate cores of O, Ne and Mg, which are susceptible to electron capture, leading to core collapse.
For KSN~2015K, an example of FELTs, \cite{2018NatAs...2..307R} prefers a CSI model, while \cite{2019ApJ...881...35T} has found that the collapse of an ONeMg star surrounded by an optically thick CSM can also explain the fast rise of the light curve.
However, the progenitors of ECSNe are thought to be super-AGB stars, which have stellar winds with relatively low velocities ($\sim$10~km~s$^{-1}$). According to theoretical predictions, ECSNe are usually faint and have low explosion energies (e.g., \citealp{2010Natur.465..326K, 2009MNRAS.398.1041B} and a most recent study \citealp{2020arXiv201102176H}).
Thus, the ECSN scenario is unlikely for AT2018cow.

\section{Summary}\label{sec:summary}
In this paper, we present our photometric and spectroscopic observations on the peculiar transient AT2018cow.
The multi-band photometry covers from peak to $\sim$70 days and spectroscopy ranges from 5 to $\sim$50 days after discovery.
The rapid rise ($t_\mathrm{r}\lesssim$2.9 days), luminous light curves ($M_{V,\mathrm{peak}}\sim-$20.8~mag) and fast post-peak decline make AT2018cow stand out of any other optical transients.
After a thorough analysis, we find that the light curves and color evolution show high resemblances to some SNe~Ibn.
With detailed analysis of the spectral evolution and line identifications, we find that AT2018cow shows similar properties to the interacting SNe, like SNe~IIn and SNe~Ibn.
Some broad emission lines due to HI, HeI, HeII, CIII, OI, and OIII emerge at $t\sim10$~days, with $v_{\mathrm{FWHM}}$ decreasing from $\sim11,000$~km~s$^{-1}$ to $\sim$3000~km~s$^{-1}$ at the end of our observations. At $t\sim20$~days, narrow and weak He~I lines ($v_{\mathrm{FWHM}}\sim$ 800-1000 km~s$^{-1}$) overlain on the broad lines. These emission lines are evidence of interaction between the ejecta and a H-rich CSM.
Furthermore, we spotted the site of AT2018cow after it faded away and find that it has a solar-like metallicity.
The host galaxy of AT2018cow has properties similar to those of GRBs and CCSNe, but is distinct from SLSNe and SNe~Ia. High star formation rate at the site of AT2018cow implies that AT2018cow might originate from a massive star.

Based on the interpretation of a CSI supernova, we fit the bolometric light curves with CSI+RD models.
We find that in order to produce the fast and bright early light curve of AT2018cow, the CSI model with a steady wind requires much larger mass loss rate than that derived from radio observations.
While with a dense uniform CSM shell, the CSI+RD model can make plausible fit with best-fit parameters $M_{\mathrm{ej}}\sim3.16$~\msun, $M_{\mathrm{CSM}}\sim0.04$~\msun, $M_{\mathrm{Ni}}\sim0.23$~\msun.
Such a CSM shell can be formed by eruptive mass ejection of LBVs immediately before core-collapse or common envelope ejection in binaries.
With $Z\approx Z_{\odot}$, the progenitor is less likely to have undergone PPI.
We conclude that the progenitor of AT2018cow is likely to be a less massive star in a binary system.

\acknowledgments
We acknowledge the support of the staff of the \mbox{Xinglong} 2.16~m and Lijiang 2.4~m telescope. This work is supported by the National Natural Science Foundation of China (NSFC grants 12033003, 11761141001, and 11633002), and the National Program on Key Research and Development Project (grant no. 2016YFA0400803). This work was also partially supported by the Open Project Program of the Key Laboratory of Optical Astronomy, National Astronomical Observatories, Chinese Academy of Sciences. This work is partially supported by the Scholar Program of Beijing Academy of Science and Technology (DZ:BS202002).
J.-J. Zhang is supported by the National Science Foundation of China (NSFC, grants 11403096, 11773067), the Key Research Program of the CAS (Grant NO. KJZD-EW- M06), the Youth Innovation Promotion Association of the CAS (grants 2018081), and the CAS ``Light of West China'' Program. X.L. was supported by the China Postdoctoral Science Foundation funded project (No: 2017m610866).
This work makes use of the LCO network. J.B., D.A.H., and D.H. were supported by NSF grant AST-1911225, and NASA grant 80NSSC19kf1639.
Research by S.V. is supported by NSF grants AST�C1813176 and AST-2008108.
We thank the staff of AZT for their observations and allowance of the use of the data.

JV and his group at Konkoly Observatory is supported by the project
``Transient Astrophysical Objects'' GINOP 2.3.2-15-2016-00033 of the
National Research, Development and Innovation Office (NKFIH), Hungary,
funded by the European Union. LK received support from the Hungarian
National Research, Development and Innovation Office grant OTKA K-131508,
and from the J\'anos Bolyai Research Scholarship of the Hungarian Academy
of Sciences. AB has been supported by the Lend\"ulet Program of the
Hungarian Academy of Sciences, project No. LP2018-7/2019.
ZsB acknowledges the support provided by the Hungarian National Research,
Development and Innovation Office (NKFIH), project No. PD-123910,
and the support by the J\'anos Bolyai Research Scholarship of the
Hungarian Academy of Sciences.

\software{ZrutyPhot (Mo et al. in prep.), IRAF \citep{1993ASPC...52..173T, 1986SPIE..627..733T}, Firefly \citep{2017MNRAS.472.4297W}.}
\bibliography{ref}

\begin{thebibliography}{}
\expandafter\ifx\csname natexlab\endcsname\relax\def\natexlab#1{#1}\fi
\providecommand{\url}[1]{\href{#1}{#1}}

\bibitem[{{Abolfathi} {et~al.}(2018){Abolfathi}, {Aguado}, {Aguilar}, {Allende
  Prieto}, {Almeida}, {Ananna}, {Anders}, {Anderson}, {Andrews}, {Anguiano},
  {Arag{\'o}n-Salamanca}, {Argudo-Fern{\'a}ndez}, {Armengaud}, {Ata},
  {Aubourg}, {Avila-Reese}, {Badenes}, {Bailey}, {Balland}, {Barger},
  {Barrera-Ballesteros}, {Bartosz}, {Bastien}, {Bates}, {Baumgarten},
  {Bautista}, {Beaton}, {Beers}, {Belfiore}, {Bender}, {Bernardi}, {Bershady},
  {Beutler}, {Bird}, {Bizyaev}, {Blanc}, {Blanton}, {Blomqvist}, {Bolton},
  {Boquien}, {Borissova}, {Bovy}, {Bradna Diaz}, {Brandt}, {Brinkmann},
  {Brownstein}, {Bundy}, {Burgasser}, {Burtin}, {Busca}, {Ca{\~n}as},
  {Cano-D{\'\i}az}, {Cappellari}, {Carrera}, {Casey}, {Cervantes Sodi}, {Chen},
  {Cherinka}, {Chiappini}, {Choi}, {Chojnowski}, {Chuang}, {Chung}, {Clerc},
  {Cohen}, {Comerford}, {Comparat}, {Correa do Nascimento}, {da Costa},
  {Cousinou}, {Covey}, {Crane}, {Cruz-Gonzalez}, {Cunha}, {da Silva Ilha},
  {Damke}, {Darling}, {Davidson}, {Dawson}, {de Icaza Lizaola}, {de la
  Macorra}, {de la Torre}, {De Lee}, {de Sainte Agathe}, {Deconto Machado},
  {Dell'Agli}, {Delubac}, {Diamond-Stanic}, {Donor}, {Downes}, {Drory}, {du Mas
  des Bourboux}, {Duckworth}, {Dwelly}, {Dyer}, {Ebelke}, {Davis Eigenbrot},
  {Eisenstein}, {Elsworth}, {Emsellem}, {Eracleous}, {Erfanianfar},
  {Escoffier}, {Fan}, {Fern{\'a}ndez Alvar}, {Fernandez-Trincado}, {Fernand o
  Cirolini}, {Feuillet}, {Finoguenov}, {Fleming}, {Font-Ribera}, {Freischlad},
  {Frinchaboy}, {Fu}, {G{\'o}mez Maqueo Chew}, {Galbany}, {Garc{\'\i}a
  P{\'e}rez}, {Garcia-Dias}, {Garc{\'\i}a-Hern{\'a}ndez}, {Garma Oehmichen},
  {Gaulme}, {Gelfand }, {Gil-Mar{\'\i}n}, {Gillespie}, {Goddard}, {Gonz{\'a}lez
  Hern{\'a}ndez}, {Gonzalez-Perez}, {Grabowski}, {Green}, {Grier}, {Gueguen},
  {Guo}, {Guy}, {Hagen}, {Hall}, {Harding}, {Hasselquist}, {Hawley}, {Hayes},
  {Hearty}, {Hekker}, {Hernand ez}, {Hernandez Toledo}, {Hogg},
  {Holley-Bockelmann}, {Holtzman}, {Hou}, {Hsieh}, {Hunt}, {Hutchinson},
  {Hwang}, {Jimenez Angel}, {Johnson}, {Jones}, {J{\"o}nsson}, {Jullo}, {Khan},
  {Kinemuchi}, {Kirkby}, {Kirkpatrick}, {Kitaura}, {Knapp}, {Kneib},
  {Kollmeier}, {Lacerna}, {Lane}, {Lang}, {Law}, {Le Goff}, {Lee}, {Li}, {Li},
  {Lian}, {Liang}, {Lima}, {Lin}, {Long}, {Lucatello}, {Lundgren}, {Mackereth},
  {MacLeod}, {Mahadevan}, {Maia}, {Majewski}, {Manchado}, {Maraston},
  {Mariappan}, {Marques-Chaves}, {Masseron}, {Masters}, {McDermid}, {McGreer},
  {Melendez}, {Meneses-Goytia}, {Merloni}, {Merrifield}, {Meszaros}, {Meza},
  {Minchev}, {Minniti}, {Mueller}, {Muller-Sanchez}, {Muna}, {Mu{\~n}oz},
  {Myers}, {Nair}, {Nand ra}, {Ness}, {Newman}, {Nichol}, {Nidever},
  {Nitschelm}, {Noterdaeme}, {O'Connell}, {Oelkers}, {Oravetz}, {Oravetz},
  {Ort{\'\i}z}, {Osorio}, {Pace}, {Padilla}, {Palanque-Delabrouille},
  {Palicio}, {Pan}, {Pan}, {Parikh}, {P{\^a}ris}, {Park}, {Peirani},
  {Pellejero-Ibanez}, {Penny}, {Percival}, {Perez-Fournon}, {Petitjean},
  {Pieri}, {Pinsonneault}, {Pisani}, {Prada}, {Prakash}, {Queiroz}, {Raddick},
  {Raichoor}, {Barboza Rembold}, {Richstein}, {Riffel}, {Riffel}, {Rix},
  {Robin}, {Rodr{\'\i}guez Torres}, {Rom{\'a}n-Z{\'u}{\~n}iga}, {Ross},
  {Rossi}, {Ruan}, {Ruggeri}, {Ruiz}, {Salvato}, {S{\'a}nchez}, {S{\'a}nchez},
  {Sanchez Almeida}, {S{\'a}nchez-Gallego}, {Santana Rojas}, {Santiago},
  {Schiavon}, {Schimoia}, {Schlafly}, {Schlegel}, {Schneider}, {Schuster},
  {Schwope}, {Seo}, {Serenelli}, {Shen}, {Shen}, {Shetrone}, {Shull}, {Silva
  Aguirre}, {Simon}, {Skrutskie}, {Slosar}, {Smethurst}, {Smith}, {Sobeck},
  {Somers}, {Souter}, {Souto}, {Spindler}, {Stark}, {Stassun}, {Steinmetz},
  {Stello}, {Storchi-Bergmann}, {Streblyanska}, {Stringfellow}, {Su{\'a}rez},
  {Sun}, {Szigeti}, {Taghizadeh-Popp}, {Talbot}, {Tang}, {Tao}, {Tayar},
  {Tembe}, {Teske}, {Thakar}, {Thomas}, {Tissera}, {Tojeiro}, {Tremonti},
  {Troup}, {Urry}, {Valenzuela}, {van den Bosch}, {Vargas-Gonz{\'a}lez},
  {Vargas-Maga{\~n}a}, {Vazquez}, {Villanova}, {Vogt}, {Wake}, {Wang},
  {Weaver}, {Weijmans}, {Weinberg}, {Westfall}, {Whelan}, {Wilcots}, {Wild},
  {Williams}, {Wilson}, {Wood-Vasey}, {Wylezalek}, {Xiao}, {Yan}, {Yang},
  {Ybarra}, {Y{\`e}che}, {Zakamska}, {Zamora}, {Zarrouk}, {Zasowski}, {Zhang},
  {Zhao}, {Zhao}, {Zheng}, {Zheng}, {Zhou}, {Zhu}, {Zinn}, \&
  {Zou}}]{2018ApJS..235...42A}
{Abolfathi}, B., {Aguado}, D.~S., {Aguilar}, G., {et~al.} 2018, \apjs, 235, 42

\bibitem[{{Angus} {et~al.}(2016){Angus}, {Levan}, {Perley}, {Tanvir}, {Lyman},
  {Stanway}, \& {Fruchter}}]{2016MNRAS.458...84A}
{Angus}, C.~R., {Levan}, A.~J., {Perley}, D.~A., {et~al.} 2016, \mnras, 458, 84

\bibitem[{{Arcavi} {et~al.}(2016){Arcavi}, {Wolf}, {Howell}, {Bildsten},
  {Leloudas}, {Hardin}, {Prajs}, {Perley}, {Svirski}, {Gal-Yam}, {Katz},
  {McCully}, {Cenko}, {Lidman}, {Sullivan}, {Valenti}, {Astier}, {Balland},
  {Carlberg}, {Conley}, {Fouchez}, {Guy}, {Pain}, {Palanque-Delabrouille},
  {Perrett}, {Pritchet}, {Regnault}, {Rich}, \&
  {Ruhlmann-Kleider}}]{2016ApJ...819...35A}
{Arcavi}, I., {Wolf}, W.~M., {Howell}, D.~A., {et~al.} 2016, \apj, 819, 35

\bibitem[{Arnett(1982)}]{1982ApJ...253..785A}
Arnett, W.~D. 1982, Astrophysical Journal, 253, 785

\bibitem[{{Bianco} {et~al.}(2014){Bianco}, {Modjaz}, {Hicken}, {Friedman},
  {Kirshner}, {Bloom}, {Challis}, {Marion}, {Wood-Vasey}, \&
  {Rest}}]{2014ApJS..213...19B}
{Bianco}, F.~B., {Modjaz}, M., {Hicken}, M., {et~al.} 2014, \apjs, 213, 19

\bibitem[{{Botticella} {et~al.}(2009){Botticella}, {Pastorello}, {Smartt},
  {Meikle}, {Benetti}, {Kotak}, {Cappellaro}, {Crockett}, {Mattila}, {Sereno},
  {Patat}, {Tsvetkov}, {van Loon}, {Abraham}, {Agnoletto}, {Arbour}, {Benn},
  {di Rico}, {Elias-Rosa}, {Gorshanov}, {Harutyunyan}, {Hunter}, {Lorenzi},
  {Keenan}, {Maguire}, {Mendez}, {Mobberley}, {Navasardyan}, {Ries},
  {Stanishev}, {Taubenberger}, {Trundle}, {Turatto}, \&
  {Volkov}}]{2009MNRAS.398.1041B}
{Botticella}, M.~T., {Pastorello}, A., {Smartt}, S.~J., {et~al.} 2009, \mnras,
  398, 1041

\bibitem[{{Brown} {et~al.}(2014){Brown}, {Breeveld}, {Holland}, {Kuin}, \&
  {Pritchard}}]{2014Ap&SS.354...89B}
{Brown}, P.~J., {Breeveld}, A.~A., {Holland}, S., {Kuin}, P., \& {Pritchard},
  T. 2014, \apss, 354, 89

\bibitem[{{Chatzopoulos} {et~al.}(2012){Chatzopoulos}, {Wheeler}, \&
  {Vinko}}]{2012ApJ...746..121C}
{Chatzopoulos}, E., {Wheeler}, J.~C., \& {Vinko}, J. 2012, \apj, 746, 121

\bibitem[{{Chatzopoulos} {et~al.}(2013){Chatzopoulos}, {Wheeler}, {Vinko},
  {Horvath}, \& {Nagy}}]{2013ApJ...773...76C}
{Chatzopoulos}, E., {Wheeler}, J.~C., {Vinko}, J., {Horvath}, Z.~L., \& {Nagy},
  A. 2013, \apj, 773, 76

\bibitem[{{Chen} {et~al.}(2014){Chen}, {Wang}, {Ganeshalingam}, {Silverman},
  {Filippenko}, {Li}, {Chornock}, {Li}, \& {Steele}}]{2014ApJ...790..120C}
{Chen}, J., {Wang}, X., {Ganeshalingam}, M., {et~al.} 2014, \apj, 790, 120

\bibitem[{{Chevalier} \& {Liang}(1989)}]{1989ApJ...344..332C}
{Chevalier}, R.~A., \& {Liang}, E.~P. 1989, \apj, 344, 332

\bibitem[{{Chugai}(2009)}]{2009MNRAS.400..866C}
{Chugai}, N.~N. 2009, \mnras, 400, 866

\bibitem[{{Curti} {et~al.}(2017){Curti}, {Cresci}, {Mannucci}, {Marconi},
  {Maiolino}, \& {Esposito}}]{2017MNRAS.465.1384C}
{Curti}, M., {Cresci}, G., {Mannucci}, F., {et~al.} 2017, \mnras, 465, 1384

\bibitem[{{Drout} {et~al.}(2011){Drout}, {Soderberg}, {Gal-Yam}, {Cenko},
  {Fox}, {Leonard}, {Sand}, {Moon}, {Arcavi}, \& {Green}}]{2011ApJ...741...97D}
{Drout}, M.~R., {Soderberg}, A.~M., {Gal-Yam}, A., {et~al.} 2011, \apj, 741, 97

\bibitem[{{Drout} {et~al.}(2014){Drout}, {Chornock}, {Soderberg}, {Sand ers},
  {McKinnon}, {Rest}, {Foley}, {Milisavljevic}, {Margutti}, {Berger},
  {Calkins}, {Fong}, {Gezari}, {Huber}, {Kankare}, {Kirshner}, {Leibler},
  {Lunnan}, {Mattila}, {Marion}, {Narayan}, {Riess}, {Roth}, {Scolnic},
  {Smartt}, {Tonry}, {Burgett}, {Chambers}, {Hodapp}, {Jedicke}, {Kaiser},
  {Magnier}, {Metcalfe}, {Morgan}, {Price}, \& {Waters}}]{2014ApJ...794...23D}
{Drout}, M.~R., {Chornock}, R., {Soderberg}, A.~M., {et~al.} 2014, \apj, 794,
  23

\bibitem[{{Dwarkadas}(2011)}]{2011MNRAS.412.1639D}
{Dwarkadas}, V.~V. 2011, \mnras, 412, 1639

\bibitem[{{Ehgamberdiev}(2018)}]{2018NatAs...2..349E}
{Ehgamberdiev}, S. 2018, Nature Astronomy, 2, 349

\bibitem[{{Eldridge} {et~al.}(2017){Eldridge}, {Stanway}, {Xiao}, {McClelland},
  {Taylor}, {Ng}, {Greis}, \& {Bray}}]{2017PASA...34...58E}
{Eldridge}, J.~J., {Stanway}, E.~R., {Xiao}, L., {et~al.} 2017, \pasa, 34, e058

\bibitem[{{Foley} {et~al.}(2007){Foley}, {Smith}, {Ganeshalingam}, {Li},
  {Chornock}, \& {Filippenko}}]{2007ApJ...657L.105F}
{Foley}, R.~J., {Smith}, N., {Ganeshalingam}, M., {et~al.} 2007, \apjl, 657,
  L105

\bibitem[{{Foley} {et~al.}(2003){Foley}, {Papenkova}, {Swift}, {Filippenko},
  {Li}, {Mazzali}, {Chornock}, {Leonard}, \& {Van Dyk}}]{2003PASP..115.1220F}
{Foley}, R.~J., {Papenkova}, M.~S., {Swift}, B.~J., {et~al.} 2003, \pasp, 115,
  1220

\bibitem[{{Fox} \& {Smith}(2019)}]{2019MNRAS.488.3772F}
{Fox}, O.~D., \& {Smith}, N. 2019, \mnras, 488, 3772

\bibitem[{{Fremling}(2018)}]{2018ATel11738....1F}
{Fremling}, C. 2018, The Astronomer's Telegram, 11738, 1

\bibitem[{{Gal-Yam} {et~al.}(2007){Gal-Yam}, {Leonard}, {Fox}, {Cenko},
  {Soderberg}, {Moon}, {Sand}, {Caltech Core Collapse Program}, {Li},
  {Filippenko}, {Aldering}, \& {Copin}}]{2007ApJ...656..372G}
{Gal-Yam}, A., {Leonard}, D.~C., {Fox}, D.~B., {et~al.} 2007, \apj, 656, 372

\bibitem[{{Galama} {et~al.}(1998){Galama}, {Vreeswijk}, {van Paradijs},
  {Kouveliotou}, {Augusteijn}, {B{\"o}hnhardt}, {Brewer}, {Doublier},
  {Gonzalez}, {Leibundgut}, {Lidman}, {Hainaut}, {Patat}, {Heise}, {in't Zand},
  {Hurley}, {Groot}, {Strom}, {Mazzali}, {Iwamoto}, {Nomoto}, {Umeda},
  {Nakamura}, {Young}, {Suzuki}, {Shigeyama}, {Koshut}, {Kippen}, {Robinson},
  {de Wildt}, {Wijers}, {Tanvir}, {Greiner}, {Pian}, {Palazzi}, {Frontera},
  {Masetti}, {Nicastro}, {Feroci}, {Costa}, {Piro}, {Peterson}, {Tinney},
  {Boyle}, {Cannon}, {Stathakis}, {Sadler}, {Begam}, \&
  {Ianna}}]{1998Natur.395..670G}
{Galama}, T.~J., {Vreeswijk}, P.~M., {van Paradijs}, J., {et~al.} 1998, \nat,
  395, 670

\bibitem[{{Galbany} {et~al.}(2014){Galbany}, {Stanishev}, {Mour{\~a}o},
  {Rodrigues}, {Flores}, {Garc{\'\i}a-Benito}, {Mast}, {Mendoza},
  {S{\'a}nchez}, {Badenes}, {Barrera-Ballesteros}, {Bland-Hawthorn},
  {Falc{\'o}n-Barroso}, {Garc{\'\i}a-Lorenzo}, {Gomes}, {Gonz{\'a}lez Delgado},
  {Kehrig}, {Lyubenova}, {L{\'o}pez-S{\'a}nchez}, {de Lorenzo-C{\'a}ceres},
  {Marino}, {Meidt}, {Moll{\'a}}, {Papaderos}, {P{\'e}rez-Torres},
  {Rosales-Ortega}, \& {van de Ven}}]{2014A&A...572A..38G}
{Galbany}, L., {Stanishev}, V., {Mour{\~a}o}, A.~M., {et~al.} 2014, \aap, 572,
  A38

\bibitem[{{Georgy} {et~al.}(2012){Georgy}, {Ekstr{\"o}m}, {Meynet}, {Massey},
  {Levesque}, {Hirschi}, {Eggenberger}, \& {Maeder}}]{2012A&A...542A..29G}
{Georgy}, C., {Ekstr{\"o}m}, S., {Meynet}, G., {et~al.} 2012, \aap, 542, A29

\bibitem[{{Guillochon} {et~al.}(2017){Guillochon}, {Parrent}, {Kelley}, \&
  {Margutti}}]{2017ApJ...835...64G}
{Guillochon}, J., {Parrent}, J., {Kelley}, L.~Z., \& {Margutti}, R. 2017, \apj,
  835, 64

\bibitem[{{Hicken} {et~al.}(2017){Hicken}, {Friedman}, {Blondin}, {Challis},
  {Berlind}, {Calkins}, {Esquerdo}, {Matheson}, {Modjaz}, {Rest}, \&
  {Kirshner}}]{2017ApJS..233....6H}
{Hicken}, M., {Friedman}, A.~S., {Blondin}, S., {et~al.} 2017, \apjs, 233, 6

\bibitem[{{Hiramatsu} {et~al.}(2020){Hiramatsu}, {Howell}, {Van Dyk},
  {Goldberg}, {Maeda}, {Moriya}, {Tominaga}, {Nomoto}, {Hosseinzadeh},
  {Arcavi}, {McCully}, {Burke}, {Bostroem}, {Valenti}, {Dong}, {Brown},
  {Andrews}, {Bilinski}, {Williams}, {Smith}, {Smith}, {Sand}, {Anand}, {Xu},
  {Filippenko}, {Bersten}, {Folatelli}, {Kelly}, {Noguchi}, \&
  {Itagaki}}]{2020arXiv201102176H}
{Hiramatsu}, D., {Howell}, D.~A., {Van Dyk}, S.~D., {et~al.} 2020, arXiv
  e-prints, arXiv:2011.02176

\bibitem[{{Ho} {et~al.}(2019){Ho}, {Phinney}, {Ravi}, {Kulkarni}, {Petitpas},
  {Emonts}, {Bhalerao}, {Blundell}, {Cenko}, {Dobie}, {Howie}, {Kamraj},
  {Kasliwal}, {Murphy}, {Perley}, {Sridharan}, \& {Yoon}}]{2019ApJ...871...73H}
{Ho}, A. Y.~Q., {Phinney}, E.~S., {Ravi}, V., {et~al.} 2019, \apj, 871, 73

\bibitem[{{Hosseinzadeh} {et~al.}(2017){Hosseinzadeh}, {Arcavi}, {Valenti},
  {McCully}, {Howell}, {Johansson}, {Sollerman}, {Pastorello}, {Benetti},
  {Cao}, {Cenko}, {Clubb}, {Corsi}, {Duggan}, {Elias-Rosa}, {Filippenko},
  {Fox}, {Fremling}, {Horesh}, {Karamehmetoglu}, {Kasliwal}, {Marion}, {Ofek},
  {Sand}, {Taddia}, {Zheng}, {Fraser}, {Gal-Yam}, {Inserra}, {Laher}, {Masci},
  {Rebbapragada}, {Smartt}, {Smith}, {Sullivan}, {Surace}, \&
  {Wo{\'z}niak}}]{2017ApJ...836..158H}
{Hosseinzadeh}, G., {Arcavi}, I., {Valenti}, S., {et~al.} 2017, \apj, 836, 158

\bibitem[{{Howell}(2017)}]{2017hsn..book..431H}
{Howell}, D.~A. 2017, {Superluminous Supernovae}, ed. A.~W. {Alsabti} \&
  P.~{Murdin}, 431

\bibitem[{{Huang} {et~al.}(2012){Huang}, {Li}, {Wang}, {Shang}, {Zhang}, {Hu},
  {Qiu}, \& {Jiang}}]{2012RAA....12.1585H}
{Huang}, F., {Li}, J.-Z., {Wang}, X.-F., {et~al.} 2012, Research in Astronomy
  and Astrophysics, 12, 1585

\bibitem[{{Iwamoto} {et~al.}(1998){Iwamoto}, {Mazzali}, {Nomoto}, {Umeda},
  {Nakamura}, {Patat}, {Danziger}, {Young}, {Suzuki}, {Shigeyama},
  {Augusteijn}, {Doublier}, {Gonzalez}, {Boehnhardt}, {Brewer}, {Hainaut},
  {Lidman}, {Leibundgut}, {Cappellaro}, {Turatto}, {Galama}, {Vreeswijk},
  {Kouveliotou}, {van Paradijs}, {Pian}, {Palazzi}, \&
  {Frontera}}]{1998Natur.395..672I}
{Iwamoto}, K., {Mazzali}, P.~A., {Nomoto}, K., {et~al.} 1998, \nat, 395, 672

\bibitem[{{Kawabata} {et~al.}(2010){Kawabata}, {Maeda}, {Nomoto},
  {Taubenberger}, {Tanaka}, {Deng}, {Pian}, {Hattori}, \&
  {Itagaki}}]{2010Natur.465..326K}
{Kawabata}, K.~S., {Maeda}, K., {Nomoto}, K., {et~al.} 2010, \nat, 465, 326

\bibitem[{{Kennicutt}(1998)}]{1998ARA&A..36..189K}
{Kennicutt}, Robert~C., J. 1998, \araa, 36, 189

\bibitem[{{Kiewe} {et~al.}(2012){Kiewe}, {Gal-Yam}, {Arcavi}, {Leonard},
  {Emilio Enriquez}, {Cenko}, {Fox}, {Moon}, {Sand }, {Soderberg}, \&
  {CCCP}}]{2012ApJ...744...10K}
{Kiewe}, M., {Gal-Yam}, A., {Arcavi}, I., {et~al.} 2012, \apj, 744, 10

\bibitem[{{Kroupa}(2001)}]{2001MNRAS.322..231K}
{Kroupa}, P. 2001, \mnras, 322, 231

\bibitem[{{Kuin} {et~al.}(2019){Kuin}, {Wu}, {Oates}, {Lien}, {Emery},
  {Kennea}, {de Pasquale}, {Han}, {Brown}, {Tohuvavohu}, {Breeveld}, {Burrows},
  {Cenko}, {Campana}, {Levan}, {Markwardt}, {Osborne}, {Page}, {Page},
  {Sbarufatti}, {Siegel}, \& {Troja}}]{2019MNRAS.487.2505K}
{Kuin}, N. P.~M., {Wu}, K., {Oates}, S., {et~al.} 2019, \mnras, 487, 2505

\bibitem[{{Leung} {et~al.}(2020){Leung}, {Blinnikov}, {Nomoto}, {Baklanov},
  {Sorokina}, \& {Tolstov}}]{2020ApJ...903...66L}
{Leung}, S.-C., {Blinnikov}, S., {Nomoto}, K., {et~al.} 2020, \apj, 903, 66

\bibitem[{{Leung} {et~al.}(2019){Leung}, {Nomoto}, \&
  {Blinnikov}}]{2019ApJ...887...72L}
{Leung}, S.-C., {Nomoto}, K., \& {Blinnikov}, S. 2019, \apj, 887, 72

\bibitem[{{Lin} {et~al.}(2020){Lin}, {Wang}, {Wang}, \&
  {Dai}}]{2020ApJ...903L..24L}
{Lin}, W.~L., {Wang}, X.~F., {Wang}, L.~J., \& {Dai}, Z.~G. 2020, \apjl, 903,
  L24

\bibitem[{{Lyman} {et~al.}(2016){Lyman}, {Bersier}, {James}, {Mazzali},
  {Eldridge}, {Fraser}, \& {Pian}}]{2016MNRAS.457..328L}
{Lyman}, J.~D., {Bersier}, D., {James}, P.~A., {et~al.} 2016, \mnras, 457, 328

\bibitem[{{Lyman} {et~al.}(2020){Lyman}, {Galbany}, {S{\'a}nchez}, {Anderson},
  {Kuncarayakti}, \& {Prieto}}]{2020MNRAS.495..992L}
{Lyman}, J.~D., {Galbany}, L., {S{\'a}nchez}, S.~F., {et~al.} 2020, \mnras,
  495, 992

\bibitem[{{Lyutikov} \& {Toonen}(2019)}]{2019MNRAS.487.5618L}
{Lyutikov}, M., \& {Toonen}, S. 2019, \mnras, 487, 5618

\bibitem[{{Maraston} \& {Str{\"o}mb{\"a}ck}(2011)}]{2011MNRAS.418.2785M}
{Maraston}, C., \& {Str{\"o}mb{\"a}ck}, G. 2011, \mnras, 418, 2785

\bibitem[{{Margutti} {et~al.}(2019){Margutti}, {Metzger}, {Chornock}, {Vurm},
  {Roth}, {Grefenstette}, {Savchenko}, {Cartier}, {Steiner}, {Terreran},
  {Margalit}, {Migliori}, {Milisavljevic}, {Alexand er}, {Bietenholz},
  {Blanchard}, {Bozzo}, {Brethauer}, {Chilingarian}, {Coppejans}, {Ducci},
  {Ferrigno}, {Fong}, {G{\"o}tz}, {Guidorzi}, {Hajela}, {Hurley}, {Kuulkers},
  {Laurent}, {Mereghetti}, {Nicholl}, {Patnaude}, {Ubertini}, {Banovetz},
  {Bartel}, {Berger}, {Coughlin}, {Eftekhari}, {Frederiks}, {Kozlova},
  {Laskar}, {Svinkin}, {Drout}, {MacFadyen}, \&
  {Paterson}}]{2019ApJ...872...18M}
{Margutti}, R., {Metzger}, B.~D., {Chornock}, R., {et~al.} 2019, \apj, 872, 18

\bibitem[{{Mauerhan} {et~al.}(2013){Mauerhan}, {Smith}, {Filippenko},
  {Blanchard}, {Blanchard}, {Casper}, {Cenko}, {Clubb}, {Cohen}, {Fuller},
  {Li}, \& {Silverman}}]{2013MNRAS.430.1801M}
{Mauerhan}, J.~C., {Smith}, N., {Filippenko}, A.~V., {et~al.} 2013, \mnras,
  430, 1801

\bibitem[{{Maund} {et~al.}(2016){Maund}, {Pastorello}, {Mattila}, {Itagaki}, \&
  {Boles}}]{2016ApJ...833..128M}
{Maund}, J.~R., {Pastorello}, A., {Mattila}, S., {Itagaki}, K., \& {Boles}, T.
  2016, \apj, 833, 128

\bibitem[{{Mazzali} {et~al.}(2008){Mazzali}, {Valenti}, {Della Valle},
  {Chincarini}, {Sauer}, {Benetti}, {Pian}, {Piran}, {D'Elia}, {Elias-Rosa},
  {Margutti}, {Pasotti}, {Antonelli}, {Bufano}, {Campana}, {Cappellaro},
  {Covino}, {D'Avanzo}, {Fiore}, {Fugazza}, {Gilmozzi}, {Hunter}, {Maguire},
  {Maiorano}, {Marziani}, {Masetti}, {Mirabel}, {Navasardyan}, {Nomoto},
  {Palazzi}, {Pastorello}, {Panagia}, {Pellizza}, {Sari}, {Smartt},
  {Tagliaferri}, {Tanaka}, {Taubenberger}, {Tominaga}, {Trundle}, \&
  {Turatto}}]{2008Sci...321.1185M}
{Mazzali}, P.~A., {Valenti}, S., {Della Valle}, M., {et~al.} 2008, Science,
  321, 1185

\bibitem[{{McKenzie} \& {Schaefer}(1999)}]{1999PASP..111..964M}
{McKenzie}, E.~H., \& {Schaefer}, B.~E. 1999, \pasp, 111, 964

\bibitem[{{Modjaz} {et~al.}(2009){Modjaz}, {Li}, {Butler}, {Chornock},
  {Perley}, {Blondin}, {Bloom}, {Filippenko}, {Kirshner}, {Kocevski},
  {Poznanski}, {Hicken}, {Foley}, {Stringfellow}, {Berlind}, {Barrado y
  Navascues}, {Blake}, {Bouy}, {Brown}, {Challis}, {Chen}, {de Vries},
  {Dufour}, {Falco}, {Friedman}, {Ganeshalingam}, {Garnavich}, {Holden},
  {Illingworth}, {Lee}, {Liebert}, {Marion}, {Olivier}, {Prochaska},
  {Silverman}, {Smith}, {Starr}, {Steele}, {Stockton}, {Williams}, \&
  {Wood-Vasey}}]{2009ApJ...702..226M}
{Modjaz}, M., {Li}, W., {Butler}, N., {et~al.} 2009, \apj, 702, 226

\bibitem[{{Moriya} {et~al.}(2014{\natexlab{a}}){Moriya}, {Maeda}, {Taddia},
  {Sollerman}, {Blinnikov}, \& {Sorokina}}]{2014MNRAS.439.2917M}
{Moriya}, T.~J., {Maeda}, K., {Taddia}, F., {et~al.} 2014{\natexlab{a}},
  \mnras, 439, 2917

\bibitem[{{Moriya} {et~al.}(2014{\natexlab{b}}){Moriya}, {Tominaga}, {Langer},
  {Nomoto}, {Blinnikov}, \& {Sorokina}}]{2014A&A...569A..57M}
{Moriya}, T.~J., {Tominaga}, N., {Langer}, N., {et~al.} 2014{\natexlab{b}},
  \aap, 569, A57

\bibitem[{{Nakamura} {et~al.}(2001){Nakamura}, {Mazzali}, {Nomoto}, \&
  {Iwamoto}}]{2001ApJ...550..991N}
{Nakamura}, T., {Mazzali}, P.~A., {Nomoto}, K., \& {Iwamoto}, K. 2001, \apj,
  550, 991

\bibitem[{{Nicholl} {et~al.}(2017){Nicholl}, {Guillochon}, \&
  {Berger}}]{2017ApJ...850...55N}
{Nicholl}, M., {Guillochon}, J., \& {Berger}, E. 2017, \apj, 850, 55

\bibitem[{{Nomoto}(1984)}]{1984ApJ...277..791N}
{Nomoto}, K. 1984, \apj, 277, 791

\bibitem[{{Nomoto}(1987)}]{1987ApJ...322..206N}
---. 1987, \apj, 322, 206

\bibitem[{{Nomoto} \& {Kondo}(1991)}]{1991ApJ...367L..19N}
{Nomoto}, K., \& {Kondo}, Y. 1991, \apjl, 367, L19

\bibitem[{{Pastorello} {et~al.}(2007){Pastorello}, {Smartt}, {Mattila},
  {Eldridge}, {Young}, {Itagaki}, {Yamaoka}, {Navasardyan}, {Valenti}, {Patat},
  {Agnoletto}, {Augusteijn}, {Benetti}, {Cappellaro}, {Boles}, {Bonnet-Bidaud},
  {Botticella}, {Bufano}, {Cao}, {Deng}, {Dennefeld}, {Elias-Rosa},
  {Harutyunyan}, {Keenan}, {Iijima}, {Lorenzi}, {Mazzali}, {Meng}, {Nakano},
  {Nielsen}, {Smoker}, {Stanishev}, {Turatto}, {Xu}, \&
  {Zampieri}}]{2007Natur.447..829P}
{Pastorello}, A., {Smartt}, S.~J., {Mattila}, S., {et~al.} 2007, \nat, 447, 829

\bibitem[{{Pastorello} {et~al.}(2008{\natexlab{a}}){Pastorello}, {Mattila},
  {Zampieri}, {Della Valle}, {Smartt}, {Valenti}, {Agnoletto}, {Benetti},
  {Benn}, {Branch}, {Cappellaro}, {Dennefeld}, {Eldridge}, {Gal-Yam},
  {Harutyunyan}, {Hunter}, {Kjeldsen}, {Lipkin}, {Mazzali}, {Milne},
  {Navasardyan}, {Ofek}, {Pian}, {Shemmer}, {Spiro}, {Stathakis},
  {Taubenberger}, {Turatto}, \& {Yamaoka}}]{2008MNRAS.389..113P}
{Pastorello}, A., {Mattila}, S., {Zampieri}, L., {et~al.} 2008{\natexlab{a}},
  \mnras, 389, 113

\bibitem[{{Pastorello} {et~al.}(2008{\natexlab{b}}){Pastorello}, {Quimby},
  {Smartt}, {Mattila}, {Navasardyan}, {Crockett}, {Elias-Rosa}, {Mondol},
  {Wheeler}, \& {Young}}]{2008MNRAS.389..131P}
{Pastorello}, A., {Quimby}, R.~M., {Smartt}, S.~J., {et~al.}
  2008{\natexlab{b}}, \mnras, 389, 131

\bibitem[{{Pastorello} {et~al.}(2015{\natexlab{a}}){Pastorello}, {Tartaglia},
  {Elias-Rosa}, {Morales-Garoffolo}, {Terreran}, {Taubenberger}, {Noebauer},
  {Benetti}, {Cappellaro}, {Ciabattari}, {Dennefeld}, {Dimai}, {Ishida},
  {Harutyunyan}, {Leonini}, {Ochner}, {Sollerman}, {Taddia}, \&
  {Zaggia}}]{2015MNRAS.454.4293P}
{Pastorello}, A., {Tartaglia}, L., {Elias-Rosa}, N., {et~al.}
  2015{\natexlab{a}}, \mnras, 454, 4293

\bibitem[{{Pastorello} {et~al.}(2015{\natexlab{b}}){Pastorello}, {Benetti},
  {Brown}, {Tsvetkov}, {Inserra}, {Taubenberger}, {Tomasella}, {Fraser},
  {Rich}, {Botticella}, {Bufano}, {Cappellaro}, {Ergon}, {Gorbovskoy},
  {Harutyunyan}, {Huang}, {Kotak}, {Lipunov}, {Magill}, {Miluzio}, {Morrell},
  {Ochner}, {Smartt}, {Sollerman}, {Spiro}, {Stritzinger}, {Turatto},
  {Valenti}, {Wang}, {Wright}, {Yurkov}, {Zampieri}, \&
  {Zhang}}]{2015MNRAS.449.1921P}
{Pastorello}, A., {Benetti}, S., {Brown}, P.~J., {et~al.} 2015{\natexlab{b}},
  \mnras, 449, 1921

\bibitem[{{Pastorello} {et~al.}(2016){Pastorello}, {Wang}, {Ciabattari},
  {Bersier}, {Mazzali}, {Gao}, {Xu}, {Zhang}, {Tokuoka}, {Benetti},
  {Cappellaro}, {Elias-Rosa}, {Harutyunyan}, {Huang}, {Miluzio}, {Mo},
  {Ochner}, {Tartaglia}, {Terreran}, {Tomasella}, \&
  {Turatto}}]{2016MNRAS.456..853P}
{Pastorello}, A., {Wang}, X.~F., {Ciabattari}, F., {et~al.} 2016, \mnras, 456,
  853

\bibitem[{{Perley} {et~al.}(2019){Perley}, {Mazzali}, {Yan}, {Cenko}, {Gezari},
  {Taggart}, {Blagorodnova}, {Fremling}, {Mockler}, \&
  {Singh}}]{2019MNRAS.484.1031P}
{Perley}, D.~A., {Mazzali}, P.~A., {Yan}, L., {et~al.} 2019, \mnras, 484, 1031

\bibitem[{{Planck Collaboration} {et~al.}(2016){Planck Collaboration}, {Ade},
  {Aghanim}, {Arnaud}, {Ashdown}, {Aumont}, {Baccigalupi}, {Banday},
  {Barreiro}, {Bartlett}, {Bartolo}, {Battaner}, {Battye}, {Benabed},
  {Beno{\^\i}t}, {Benoit-L{\'e}vy}, {Bernard}, {Bersanelli}, {Bielewicz},
  {Bock}, {Bonaldi}, {Bonavera}, {Bond}, {Borrill}, {Bouchet}, {Boulanger},
  {Bucher}, {Burigana}, {Butler}, {Calabrese}, {Cardoso}, {Catalano},
  {Challinor}, {Chamballu}, {Chary}, {Chiang}, {Chluba}, {Christensen},
  {Church}, {Clements}, {Colombi}, {Colombo}, {Combet}, {Coulais}, {Crill},
  {Curto}, {Cuttaia}, {Danese}, {Davies}, {Davis}, {de Bernardis}, {de Rosa},
  {de Zotti}, {Delabrouille}, {D{\'e}sert}, {Di Valentino}, {Dickinson},
  {Diego}, {Dolag}, {Dole}, {Donzelli}, {Dor{\'e}}, {Douspis}, {Ducout},
  {Dunkley}, {Dupac}, {Efstathiou}, {Elsner}, {En{\ss}lin}, {Eriksen},
  {Farhang}, {Fergusson}, {Finelli}, {Forni}, {Frailis}, {Fraisse},
  {Franceschi}, {Frejsel}, {Galeotta}, {Galli}, {Ganga}, {Gauthier}, {Gerbino},
  {Ghosh}, {Giard}, {Giraud-H{\'e}raud}, {Giusarma}, {Gjerl{\o}w},
  {Gonz{\'a}lez-Nuevo}, {G{\'o}rski}, {Gratton}, {Gregorio}, {Gruppuso},
  {Gudmundsson}, {Hamann}, {Hansen}, {Hanson}, {Harrison}, {Helou},
  {Henrot-Versill{\'e}}, {Hern{\'a}ndez-Monteagudo}, {Herranz}, {Hildebrand t},
  {Hivon}, {Hobson}, {Holmes}, {Hornstrup}, {Hovest}, {Huang}, {Huffenberger},
  {Hurier}, {Jaffe}, {Jaffe}, {Jones}, {Juvela}, {Keih{\"a}nen}, {Keskitalo},
  {Kisner}, {Kneissl}, {Knoche}, {Knox}, {Kunz}, {Kurki-Suonio}, {Lagache},
  {L{\"a}hteenm{\"a}ki}, {Lamarre}, {Lasenby}, {Lattanzi}, {Lawrence}, {Leahy},
  {Leonardi}, {Lesgourgues}, {Levrier}, {Lewis}, {Liguori}, {Lilje},
  {Linden-V{\o}rnle}, {L{\'o}pez-Caniego}, {Lubin}, {Mac{\'\i}as-P{\'e}rez},
  {Maggio}, {Maino}, {Mandolesi}, {Mangilli}, {Marchini}, {Maris}, {Martin},
  {Martinelli}, {Mart{\'\i}nez-Gonz{\'a}lez}, {Masi}, {Matarrese}, {McGehee},
  {Meinhold}, {Melchiorri}, {Melin}, {Mendes}, {Mennella}, {Migliaccio},
  {Millea}, {Mitra}, {Miville-Desch{\^e}nes}, {Moneti}, {Montier}, {Morgante},
  {Mortlock}, {Moss}, {Munshi}, {Murphy}, {Naselsky}, {Nati}, {Natoli},
  {Netterfield}, {N{\o}rgaard-Nielsen}, {Noviello}, {Novikov}, {Novikov},
  {Oxborrow}, {Paci}, {Pagano}, {Pajot}, {Paladini}, {Paoletti}, {Partridge},
  {Pasian}, {Patanchon}, {Pearson}, {Perdereau}, {Perotto}, {Perrotta},
  {Pettorino}, {Piacentini}, {Piat}, {Pierpaoli}, {Pietrobon}, {Plaszczynski},
  {Pointecouteau}, {Polenta}, {Popa}, {Pratt}, {Pr{\'e}zeau}, {Prunet},
  {Puget}, {Rachen}, {Reach}, {Rebolo}, {Reinecke}, {Remazeilles}, {Renault},
  {Renzi}, {Ristorcelli}, {Rocha}, {Rosset}, {Rossetti}, {Roudier},
  {Rouill{\'e} d'Orfeuil}, {Rowan-Robinson}, {Rubi{\~n}o-Mart{\'\i}n},
  {Rusholme}, {Said}, {Salvatelli}, {Salvati}, {Sandri}, {Santos},
  {Savelainen}, {Savini}, {Scott}, {Seiffert}, {Serra}, {Shellard}, {Spencer},
  {Spinelli}, {Stolyarov}, {Stompor}, {Sudiwala}, {Sunyaev}, {Sutton},
  {Suur-Uski}, {Sygnet}, {Tauber}, {Terenzi}, {Toffolatti}, {Tomasi},
  {Tristram}, {Trombetti}, {Tucci}, {Tuovinen}, {T{\"u}rler}, {Umana},
  {Valenziano}, {Valiviita}, {Van Tent}, {Vielva}, {Villa}, {Wade}, {Wandelt},
  {Wehus}, {White}, {White}, {Wilkinson}, {Yvon}, {Zacchei}, \&
  {Zonca}}]{2016A&A...594A..13P}
{Planck Collaboration}, {Ade}, P.~A.~R., {Aghanim}, N., {et~al.} 2016, \aap,
  594, A13

\bibitem[{{Prentice} {et~al.}(2018){Prentice}, {Maguire}, {Smartt}, {Magee},
  {Schady}, {Sim}, {Chen}, {Clark}, {Colin}, {Fulton}, {McBrien}, {O'Neill},
  {Smith}, {Ashall}, {Chambers}, {Denneau}, {Flewelling}, {Heinze}, {Holoien},
  {Huber}, {Kochanek}, {Mazzali}, {Prieto}, {Rest}, {Shappee}, {Stalder},
  {Stanek}, {Stritzinger}, {Thompson}, \& {Tonry}}]{2018ApJ...865L...3P}
{Prentice}, S.~J., {Maguire}, K., {Smartt}, S.~J., {et~al.} 2018, \apjl, 865,
  L3

\bibitem[{{Quimby} {et~al.}(2011){Quimby}, {Kulkarni}, {Kasliwal}, {Gal-Yam},
  {Arcavi}, {Sullivan}, {Nugent}, {Thomas}, {Howell}, {Nakar}, {Bildsten},
  {Theissen}, {Law}, {Dekany}, {Rahmer}, {Hale}, {Smith}, {Ofek}, {Zolkower},
  {Velur}, {Walters}, {Henning}, {Bui}, {McKenna}, {Poznanski}, {Cenko}, \&
  {Levitan}}]{2011Natur.474..487Q}
{Quimby}, R.~M., {Kulkarni}, S.~R., {Kasliwal}, M.~M., {et~al.} 2011, \nat,
  474, 487

\bibitem[{{Rest} {et~al.}(2018){Rest}, {Garnavich}, {Khatami}, {Kasen},
  {Tucker}, {Shaya}, {Olling}, {Mushotzky}, {Zenteno}, \&
  {Margheim}}]{2018NatAs...2..307R}
{Rest}, A., {Garnavich}, P.~M., {Khatami}, D., {et~al.} 2018, Nature Astronomy,
  2, 307

\bibitem[{{Richmond} {et~al.}(1996){Richmond}, {van Dyk}, {Ho}, {Peng}, {Paik},
  {Treffers}, {Filippenko}, {Bustamante-Donas}, {Moeller}, {Pawellek},
  {Tartara}, \& {Spence}}]{1996AJ....111..327R}
{Richmond}, M.~W., {van Dyk}, S.~D., {Ho}, W., {et~al.} 1996, \aj, 111, 327

\bibitem[{{Schlafly} \& {Finkbeiner}(2011)}]{2011ApJ...737..103S}
{Schlafly}, E.~F., \& {Finkbeiner}, D.~P. 2011, \apj, 737, 103

\bibitem[{{Shivvers} {et~al.}(2016){Shivvers}, {Zheng}, {Mauerhan}, {Kleiser},
  {Van Dyk}, {Silverman}, {Graham}, {Kelly}, {Filippenko}, \&
  {Kumar}}]{2016MNRAS.461.3057S}
{Shivvers}, I., {Zheng}, W.~K., {Mauerhan}, J., {et~al.} 2016, \mnras, 461,
  3057

\bibitem[{{Smartt} {et~al.}(2015){Smartt}, {Valenti}, {Fraser}, {Inserra},
  {Young}, {Sullivan}, {Pastorello}, {Benetti}, {Gal-Yam}, {Knapic},
  {Molinaro}, {Smareglia}, {Smith}, {Taubenberger}, {Yaron}, {Anderson},
  {Ashall}, {Balland}, {Baltay}, {Barbarino}, {Bauer}, {Baumont}, {Bersier},
  {Blagorodnova}, {Bongard}, {Botticella}, {Bufano}, {Bulla}, {Cappellaro},
  {Campbell}, {Cellier-Holzem}, {Chen}, {Childress}, {Clocchiatti},
  {Contreras}, {Dall'Ora}, {Danziger}, {de Jaeger}, {De Cia}, {Della Valle},
  {Dennefeld}, {Elias-Rosa}, {Elman}, {Feindt}, {Fleury}, {Gall},
  {Gonzalez-Gaitan}, {Galbany}, {Morales Garoffolo}, {Greggio}, {Guillou},
  {Hachinger}, {Hadjiyska}, {Hage}, {Hillebrandt}, {Hodgkin}, {Hsiao}, {James},
  {Jerkstrand}, {Kangas}, {Kankare}, {Kotak}, {Kromer}, {Kuncarayakti},
  {Leloudas}, {Lundqvist}, {Lyman}, {Hook}, {Maguire}, {Manulis}, {Margheim},
  {Mattila}, {Maund}, {Mazzali}, {McCrum}, {McKinnon}, {Moreno-Raya},
  {Nicholl}, {Nugent}, {Pain}, {Pignata}, {Phillips}, {Polshaw}, {Pumo},
  {Rabinowitz}, {Reilly}, {Romero-Ca{\~n}izales}, {Scalzo}, {Schmidt},
  {Schulze}, {Sim}, {Sollerman}, {Taddia}, {Tartaglia}, {Terreran},
  {Tomasella}, {Turatto}, {Walker}, {Walton}, {Wyrzykowski}, {Yuan}, \&
  {Zampieri}}]{2015A&A...579A..40S}
{Smartt}, S.~J., {Valenti}, S., {Fraser}, M., {et~al.} 2015, \aap, 579, A40

\bibitem[{{Smartt} {et~al.}(2018){Smartt}, {Clark}, {Smith}, {McBrien},
  {Maguire}, {O'Neil}, {Fulton}, {Magee}, {Prentice}, {Colin}, {Tonry},
  {Denneau}, {Stalder}, {Heinze}, {Weiland}, {Flewelling}, \&
  {Rest}}]{2018ATel11727....1S}
{Smartt}, S.~J., {Clark}, P., {Smith}, K.~W., {et~al.} 2018, The Astronomer's
  Telegram, 11727

\bibitem[{{Smith} {et~al.}(2012{\natexlab{a}}){Smith}, {Nichol}, {Dilday},
  {Marriner}, {Kessler}, {Bassett}, {Cinabro}, {Frieman}, {Garnavich}, {Jha},
  {Lampeitl}, {Sako}, {Schneider}, \& {Sollerman}}]{2012ApJ...755...61S}
{Smith}, M., {Nichol}, R.~C., {Dilday}, B., {et~al.} 2012{\natexlab{a}}, \apj,
  755, 61

\bibitem[{{Smith}(2014)}]{2014ARA&A..52..487S}
{Smith}, N. 2014, \araa, 52, 487

\bibitem[{{Smith} {et~al.}(2008){Smith}, {Foley}, \&
  {Filippenko}}]{2008ApJ...680..568S}
{Smith}, N., {Foley}, R.~J., \& {Filippenko}, A.~V. 2008, \apj, 680, 568

\bibitem[{{Smith} {et~al.}(2012{\natexlab{b}}){Smith}, {Mauerhan}, {Silverman},
  {Ganeshalingam}, {Filippenko}, {Cenko}, {Clubb}, \& {Kand
  rashoff}}]{2012MNRAS.426.1905S}
{Smith}, N., {Mauerhan}, J.~C., {Silverman}, J.~M., {et~al.}
  2012{\natexlab{b}}, \mnras, 426, 1905

\bibitem[{{Smith} {et~al.}(2012{\natexlab{c}}){Smith}, {Silverman},
  {Filippenko}, {Cooper}, {Matheson}, {Bian}, {Weiner}, \&
  {Comerford}}]{2012AJ....143...17S}
{Smith}, N., {Silverman}, J.~M., {Filippenko}, A.~V., {et~al.}
  2012{\natexlab{c}}, \aj, 143, 17

\bibitem[{{Sollerman} {et~al.}(2000){Sollerman}, {Kozma}, {Fransson},
  {Leibundgut}, {Lundqvist}, {Ryde}, \& {Woudt}}]{2000ApJ...537L.127S}
{Sollerman}, J., {Kozma}, C., {Fransson}, C., {et~al.} 2000, \apjl, 537, L127

\bibitem[{{Sullivan} {et~al.}(2001){Sullivan}, {Mobasher}, {Chan}, {Cram},
  {Ellis}, {Treyer}, \& {Hopkins}}]{2001ApJ...558...72S}
{Sullivan}, M., {Mobasher}, B., {Chan}, B., {et~al.} 2001, \apj, 558, 72

\bibitem[{{Sun} {et~al.}(2020){Sun}, {Maund}, {Hirai}, {Crowther}, \&
  {Podsiadlowski}}]{2020MNRAS.491.6000S}
{Sun}, N.-C., {Maund}, J.~R., {Hirai}, R., {Crowther}, P.~A., \&
  {Podsiadlowski}, P. 2020, \mnras, 491, 6000

\bibitem[{{Svensson} {et~al.}(2010){Svensson}, {Levan}, {Tanvir}, {Fruchter},
  \& {Strolger}}]{2010MNRAS.405...57S}
{Svensson}, K.~M., {Levan}, A.~J., {Tanvir}, N.~R., {Fruchter}, A.~S., \&
  {Strolger}, L.~G. 2010, \mnras, 405, 57

\bibitem[{{Taddia} {et~al.}(2015){Taddia}, {Sollerman}, {Fremling},
  {Pastorello}, {Leloudas}, {Fransson}, {Nyholm}, {Stritzinger}, {Ergon},
  {Roy}, \& {Migotto}}]{2015A&A...580A.131T}
{Taddia}, F., {Sollerman}, J., {Fremling}, C., {et~al.} 2015, \aap, 580, A131

\bibitem[{{Taubenberger} {et~al.}(2006){Taubenberger}, {Pastorello}, {Mazzali},
  {Valenti}, {Pignata}, {Sauer}, {Arbey}, {B{\"a}rnbantner}, {Benetti}, {Della
  Valle}, {Deng}, {Elias-Rosa}, {Filippenko}, {Foley}, {Goobar}, {Kotak}, {Li},
  {Meikle}, {Mendez}, {Patat}, {Pian}, {Ries}, {Ruiz-Lapuente}, {Salvo},
  {Stanishev}, {Turatto}, \& {Hillebrandt}}]{2006MNRAS.371.1459T}
{Taubenberger}, S., {Pastorello}, A., {Mazzali}, P.~A., {et~al.} 2006, \mnras,
  371, 1459

\bibitem[{{Tauris} {et~al.}(2015){Tauris}, {Langer}, \&
  {Podsiadlowski}}]{2015MNRAS.451.2123T}
{Tauris}, T.~M., {Langer}, N., \& {Podsiadlowski}, P. 2015, \mnras, 451, 2123

\bibitem[{{Tody}(1986)}]{1986SPIE..627..733T}
{Tody}, D. 1986, in \procspie, Vol. 627, Instrumentation in astronomy VI, ed.
  D.~L. {Crawford}, 733

\bibitem[{{Tody}(1993)}]{1993ASPC...52..173T}
{Tody}, D. 1993, in Astronomical Society of the Pacific Conference Series,
  Vol.~52, Astronomical Data Analysis Software and Systems II, ed. R.~J.
  {Hanisch}, R.~J.~V. {Brissenden}, \& J.~{Barnes}, 173

\bibitem[{{Tolstov} {et~al.}(2019){Tolstov}, {Nomoto}, {Sorokina}, {Blinnikov},
  {Tominaga}, \& {Taniguchi}}]{2019ApJ...881...35T}
{Tolstov}, A., {Nomoto}, K., {Sorokina}, E., {et~al.} 2019, \apj, 881, 35

\bibitem[{{Tominaga} {et~al.}(2008){Tominaga}, {Limongi}, {Suzuki}, {Tanaka},
  {Nomoto}, {Maeda}, {Chieffi}, {Tornambe}, {Minezaki}, {Yoshii}, {Sakon},
  {Wada}, {Ohyama}, {Tanab{\'e}}, {Kaneda}, {Onaka}, {Nozawa}, {Kozasa},
  {Kawabata}, {Anupama}, {Sahu}, {Gurugubelli}, {Prabhu}, \&
  {Deng}}]{2008ApJ...687.1208T}
{Tominaga}, N., {Limongi}, M., {Suzuki}, T., {et~al.} 2008, \apj, 687, 1208

\bibitem[{{Tsvetkov} {et~al.}(2015){Tsvetkov}, {Volkov}, \&
  {Pavlyuk}}]{2015IBVS.6140....1T}
{Tsvetkov}, D.~Y., {Volkov}, I.~M., \& {Pavlyuk}, N.~N. 2015, Information
  Bulletin on Variable Stars, 6140, 1

\bibitem[{{Van Dyk} {et~al.}(2018){Van Dyk}, {Zheng}, {Brink}, {Filippenko},
  {Milisavljevic}, {Andrews}, {Smith}, {Cignoni}, {Fox}, {Kelly}, {Adamo},
  {Yunus}, {Zhang}, \& {Kumar}}]{2018ApJ...860...90V}
{Van Dyk}, S.~D., {Zheng}, W., {Brink}, T.~G., {et~al.} 2018, \apj, 860, 90

\bibitem[{{Wang} {et~al.}(2019){Wang}, {Wang}, {Cano}, {Wang}, {Liu}, {Dai},
  {Deng}, {Yu}, {Li}, {Song}, {Qiu}, \& {Wei}}]{2019MNRAS.489.1110W}
{Wang}, L.~J., {Wang}, X.~F., {Cano}, Z., {et~al.} 2019, \mnras, 489, 1110

\bibitem[{{Whitesides} {et~al.}(2017){Whitesides}, {Lunnan}, {Kasliwal},
  {Perley}, {Corsi}, {Cenko}, {Blagorodnova}, {Cao}, {Cook}, {Doran},
  {Frederiks}, {Fremling}, {Hurley}, {Karamehmetoglu}, {Kulkarni}, {Leloudas},
  {Masci}, {Nugent}, {Ritter}, {Rubin}, {Savchenko}, {Sollerman}, {Svinkin},
  {Taddia}, {Vreeswijk}, \& {Wozniak}}]{2017ApJ...851..107W}
{Whitesides}, L., {Lunnan}, R., {Kasliwal}, M.~M., {et~al.} 2017, \apj, 851,
  107

\bibitem[{{Wilkinson} {et~al.}(2017){Wilkinson}, {Maraston}, {Goddard},
  {Thomas}, \& {Parikh}}]{2017MNRAS.472.4297W}
{Wilkinson}, D.~M., {Maraston}, C., {Goddard}, D., {Thomas}, D., \& {Parikh},
  T. 2017, \mnras, 472, 4297

\bibitem[{{Xiang} {et~al.}(2019){Xiang}, {Wang}, {Mo}, {Wang}, {Smartt},
  {Fraser}, {Ehgamberdiev}, {Mirzaqulov}, {Zhang}, {Zhang}, {Vinko}, {Wheeler},
  {Hosseinzadeh}, {Howell}, {McCully}, {DerKacy}, {Baron}, {Brown}, {Zhang},
  {Bi}, {Song}, {Zhang}, {Rest}, {Nomoto}, {Tolstov}, \&
  {Blinnikov}}]{2019ApJ...871..176X}
{Xiang}, D., {Wang}, X., {Mo}, J., {et~al.} 2019, \apj, 871, 176

\bibitem[{{Yokoo} {et~al.}(1994){Yokoo}, {Arimoto}, {Matsumoto}, {Takahashi},
  \& {Sadakane}}]{1994PASJ...46L.191Y}
{Yokoo}, T., {Arimoto}, J., {Matsumoto}, K., {Takahashi}, A., \& {Sadakane}, K.
  1994, \pasj, 46, L191

\bibitem[{{Zhang} {et~al.}(2016){Zhang}, {Wang}, {Zhang}, {Zhang},
  {Ganeshalingam}, {Li}, {Filippenko}, {Zhao}, {Zheng}, {Bai}, {Chen}, {Chen},
  {Huang}, {Mo}, {Rui}, {Song}, {Sai}, {Li}, {Wang}, \&
  {Wu}}]{2016ApJ...820...67Z}
{Zhang}, K., {Wang}, X., {Zhang}, J., {et~al.} 2016, \apj, 820, 67

\bibitem[{{Zhang} {et~al.}(2012){Zhang}, {Wang}, {Wu}, {Chen}, {Chen}, {Liu},
  {Huang}, {Liang}, {Zhao}, {Lin}, {Wang}, {Dennefeld}, {Zhang}, {Zhai}, {Wu},
  {Fan}, {Zou}, {Zhou}, \& {Ma}}]{2012AJ....144..131Z}
{Zhang}, T., {Wang}, X., {Wu}, C., {et~al.} 2012, \aj, 144, 131

\end{thebibliography}

\end{document}